\documentclass[12pt,preprint]{emulateapj}

\newcommand{\be}{\begin{equation}}
\newcommand{\ee}{\end{equation}}
\newcommand{\ba}{\begin{eqnarray}}
\newcommand{\ea}{\end{eqnarray}}

\begin{document}
\title{Intensity mapping of H$\alpha$, H$\beta$, [OII] and [OIII] lines at $z<5$}

\author{Yan Gong$^{1,2}$, Asantha Cooray$^2$, Marta B. Silva$^3$, Michael Zemcov$^{4,5}$, Chang Feng$^2$, \\Mario G. Santos$^{6,7}$, Olivier Dore$^{5,8}$, Xuelei Chen$^{1,9,10}$}

\affil{$^{1}$ Key Laboratory for Computational Astrophysics, National Astronomical Observatories, \\Chinese Academy of Sciences, 20A Datun Road, Beijing 100012, China \\
$^2$Department of Physics \& Astronomy, University of California, Irvine, CA 92697 \\
$^3$Kapteyn Astronomical Institute, University of Groningen, Landleven 12, \\9747AD Groningen, the Netherlands \\
$^4$Center for Detectors, School of Physics and Astronomy, Rochester Institute of Technology, \\Rochester, NY 14623, USA \\
$^5$NASA Jet Propulsion Laboratory, California Institute of Technology, 4800 Oak Grove Drive, \\MS 169-215, Pasadena, CA, 91109, USA\\
$^6$Physics Department, University of the Western Cape, Cape Town 7535, South Africa \\
$^7$SKA SA, The Park, Park Road, Cape Town 7405, South Africa \\
$^{8}$California Institute of Technology, MC 249-17, Pasadena, California, 91125 USA \\
$^{9}$University of Chinese Academy of Sciences, Beijing 100049, China\\
$^{10}$Center of High Energy Physics, Peking University, Beijing 100871, China
}

\begin{abstract}
Intensity mapping is now becoming a useful tool to study the large-scale structure of the universe through spatial variations in the integrated emission from galaxies and the intergalactic medium. We study intensity mapping of the H$\alpha\,6563\rm\AA$, [OIII]$\,5007\rm\AA$, [OII]$\,3727\rm\AA$ and H$\beta\,4861\rm\AA$ lines at $0.8\le z\le5.2$. The mean intensities of these four emission lines are estimated using the observed luminosity functions (LFs), cosmological simulations, and the star formation rate density (SFRD) derived from observations at $z\lesssim5$. We calculate the intensity power spectra and consider the foreground contamination of other lines at lower redshifts. We use the proposed NASA small explorer SPHEREx (the Spectro-Photometer for the History of the Universe, Epoch of Reionization, and Ices Explorer) as a case study for the detectability of the intensity power spectra of the four emission lines. We also investigate the cross correlation with the 21-cm line probed by CHIME (the Canadian Hydrogen Intensity Mapping Experiment), Tianlai experiment and SKA (the Square Kilometer Array) at $0.8\le z\le2.4$. We find both the auto and cross power spectra can be well measured for the H$\alpha$, [OIII] and [OII] lines at $z\lesssim 3$, while it is more challenging for the H$\beta$ line. Finally, we estimate the constraint on the SFRD from intensity mapping, and find we can reach accuracy higher than $7\%$ at $z\lesssim4$, which is better than usual measurements using the LFs of galaxies.
\end{abstract}


\maketitle

\section{Introduction}

Measurements of the large-scale structure (LSS) of the Universe are essential for studies of dark matter, dark energy and other aspects of cosmology. Ordinary galaxy surveys focus on individual galaxies and map the space distribution object by object, thereby tracing the LSS of underlying dark matter. These surveys have allowed successful measurements of  cosmic LSS at low redshifts with $z\lesssim1$, e.g. the Sloan Digital Sky Surveys (SDSS)\footnote{\tt http://www.sdss.org/}. At higher redshifts, galaxies become fainter and smaller on average, making individual detections challenging. However, the key to important questions regarding dark energy, galaxy evolution, and cosmic LSS lies at $1\lesssim z\lesssim3$ and even higher redshifts. A method that does not rely on individual source detection is necessary to make progress. 

Intensity mapping of atomic and molecular lines provides a suitable tool for the cosmological study. It can probe the faint and remote galaxies with large spatial volume in acceptable observation time, which does not need to resolve individual sources. Since the emission lines of atoms and molecules are tightly coupled to the stellar content and environment of host galaxy, intensity mapping of emission lines can also provide statistical information about the star formation rate (SFR) and other galaxy properties . 

Recently, intensity mapping of atomic and molecular emission lines, for studying the epoch of reionization (EoR) and the epochs before and after EoR at high redshifts, such as CO, [CII], Ly$\alpha$, H$_2$, etc, have been discussed \citep{Visbal10,Gong11a,Carilli11,Lidz11,Gong12,Gong13,Silva13,Pullen14,Uzgil14,Gong14,Silva15,Fonseca16}. These works indicate that intensity mapping of atomic and molecular emission lines is a powerful tool for studying the high-$z$ Universe, and it complements the method of using low-frequency radio experiments that measure the 21-cm flip-spin line from neutral hydrogen.

In this work, we study intensity mapping of four optical luminous lines, i.e. H$\alpha\,6563\rm\AA$, [OIII]$\,5007\rm\AA$, [OII]$\,3727\rm\AA$ and H$\beta\,4861\rm\AA$, in the redshift range $0.8\le z\le5.2$. We investigate the mean intensities of the four lines, and estimate the intensity power spectra at different redshifts. Foreground contamination from other lines at lower redshifts is also discussed, which is an important issue in intensity mapping surveys. In order to evaluate the detectability of the power spectra in a real measurement, we study the proposed space telescope Spectro-Photometer for the History of the Universe, Epoch of Reionization, and Ices Explorer (SPHEREx)\footnote{\tt http://spherex.caltech.edu/}, and estimate the errors and signal to noise ratio (SNR) for the intensity power spectra at $0.8\le z\le5.2$. The cross correlation of these lines with the hydrogen 21-cm line is also studied at $0.8\le z\le2.4$. The cross correlation can effectively reduce the foreground line contamination and instrument noise, and offers a reliable way to extract the signal. We discuss two 21-cm experiments in our work, the Canadian Hydrogen Intensity Mapping Experiment (CHIME\footnote{\tt http://chime.phas.ubc.ca/}) and Chinese experiment ``Tianlai''\footnote{\tt http://tianlai.bao.ac.cn/}. Finally, we explore constraint on the star formation rate density (SFRD) at different redshifts available from this intensity mapping survey.

This paper is organized as follows: in Section 2, we estimate the mean intensities of the lines at $z<6$ using three different methods. In Section 3, we calculate the intensity power spectra of the four lines at $0.8\le z\le5.2$ using a halo model. In Section 4, we discuss foreground contamination and the detectability of the line intensity power spectra with the SPHEREx experiment. In Section 5, we explore the cross correlation with the 21-cm line measured by CHIME, Tianlai experiments and SKA. In Section 6, the constraint of the SFRD at different redshifts is predicted with the intensity mapping measurements. In this paper, we adopt the flat $\Lambda$CDM model with $\Omega_b=0.046$, $\Omega_M=0.27$, $\Omega_{\Lambda}=0.73$, $h=0.71$, $\sigma_8=0.81$ and $n_s=0.96$ as the fiducial model.

\section{Line mean intensity}

In this section, we estimate the observed mean line intensities of H$\alpha$, [OIII], [OII] and  H$\beta$ at $z<5$. We make use of the observed luminosity functions, the star formation rates calculated from simulations and the SFRD derived from observations to estimate the mean intensity for each line. The effect of dust extinction is also considered in our estimates.

\subsection{Mean intensity from luminosity function}

\begin{figure*}[t]
\epsscale{1.9}
\centerline{
\resizebox{!}{!}{\includegraphics[scale=0.3]{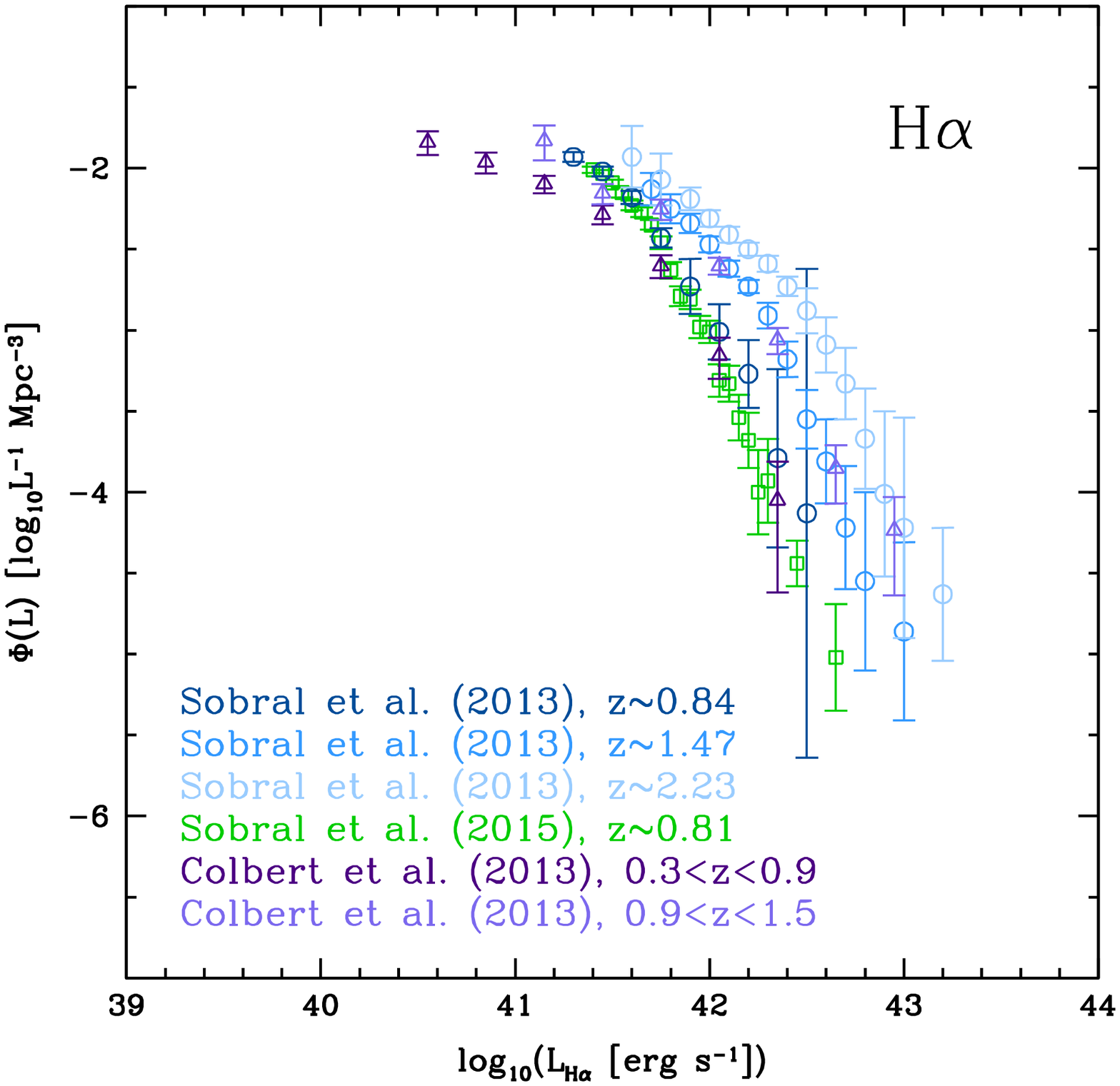}}
\resizebox{!}{!}{\includegraphics[scale=0.3]{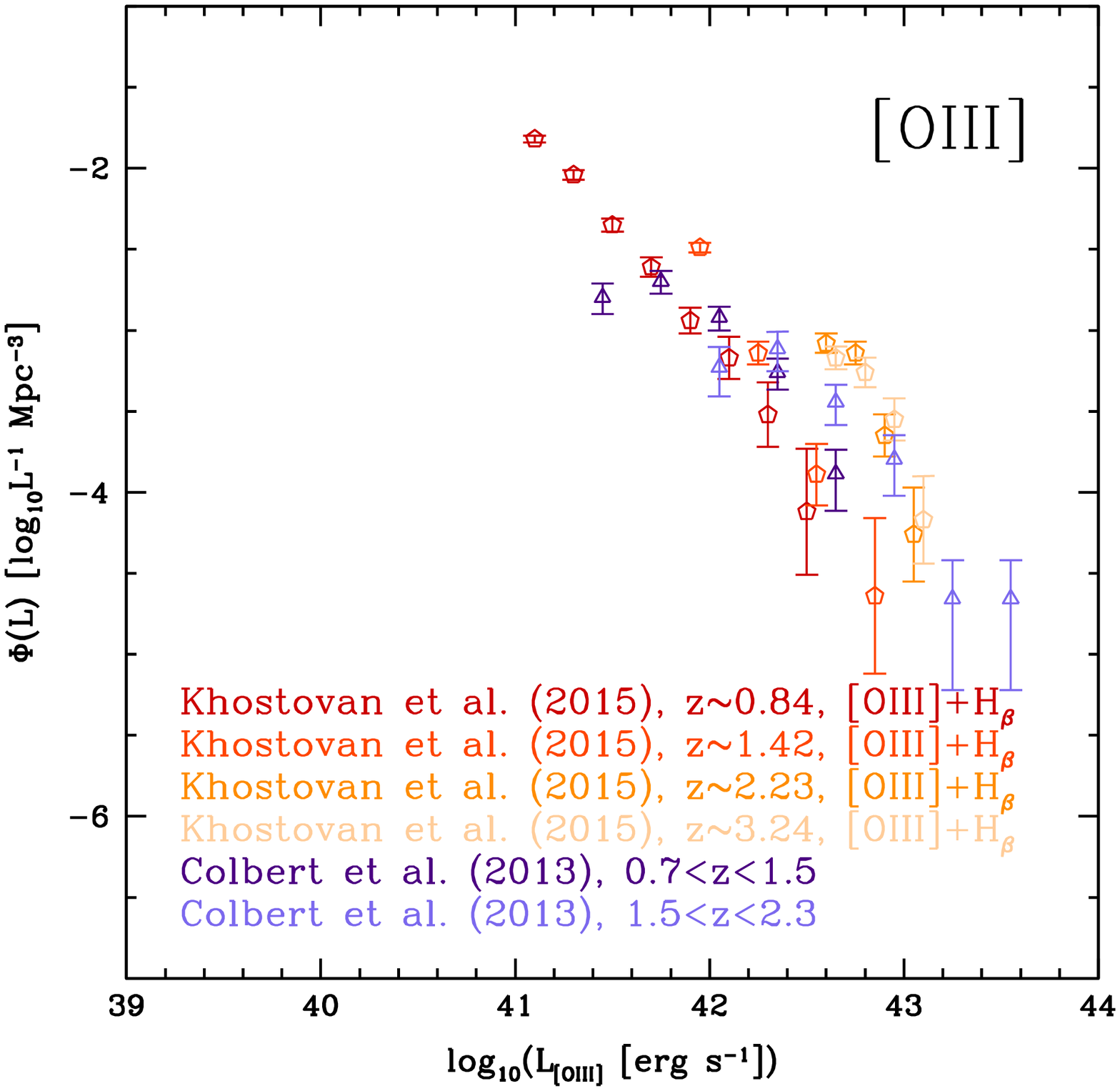}}
\resizebox{!}{!}{\includegraphics[scale=0.3]{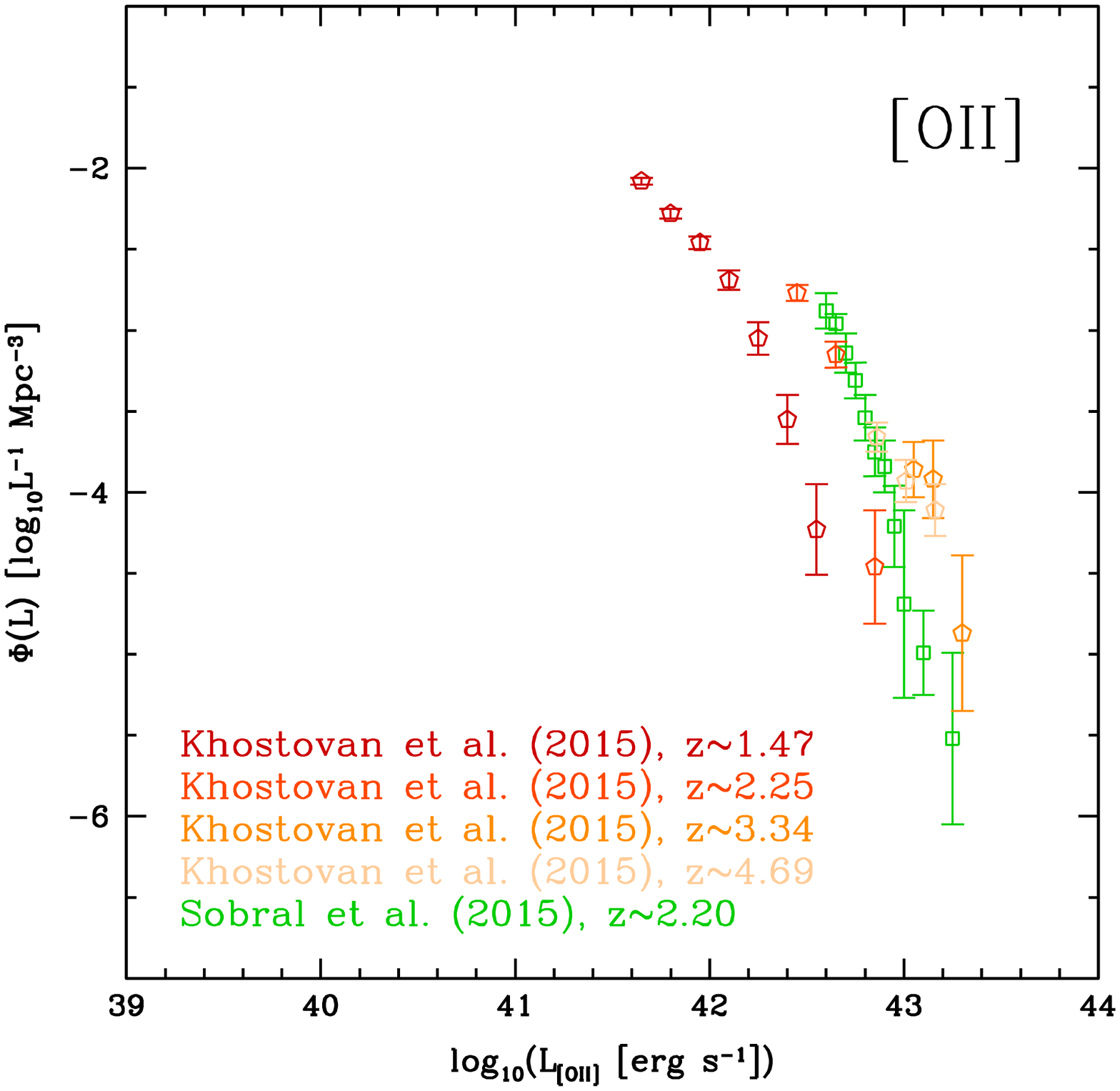}}
}
\epsscale{1.0}
\caption{\label{fig:LFs} The observed luminosity functions of H$\alpha$, [OIII] and [OII] used in the estimation of mean intensity \citep{Sobral13, Colbert13, Sobral15, Khostovan15}. These LFs are not corrected for dust extinction. The H$\alpha$ LFs in \cite{Sobral13} have had dust extinction of $A_{{\rm H}\alpha}=1$ mag reapplied to match the other data sets. The LFs in \cite{Khostovan15} are for [OIII]+H$\beta$ instead of [OIII] only.}
\end{figure*}

The direct way to estimate the mean intensity of an emission line is to use the observed line luminosity functions (LFs). In Figure \ref{fig:LFs}, we show the luminosity function data of H$\alpha$, [OIII] and [OII] at different redshifts, and we adopt their fitting formulae in our calculation \citep{Sobral13, Colbert13, Sobral15, Khostovan15}. These luminosity functions are not corrected for dust extinction, since we are trying to derive the observed mean intensity for each line. The H$\alpha$ LFs given in \cite{Sobral13} have been corrected for dust extinction with $A_{\rm H\alpha}=1$ mag, and we use the same extinction law and rescale these LFs back to including dust extinction effect. The LFs are fitted by the Schechter function \citep{Schechter76}
\be\label{eq:LF}
\Phi(L)dL = \phi_*\left( \frac{L}{L_*} \right)^{\alpha} {\rm exp}\left(-\frac{L}{L_*}\right)\frac{dL}{L_*},
\ee
where $\phi_*$, $\alpha$, and $L_*$ are the free parameters obtained by fitting observational data. The observed mean intensity can then be estimated from
\be \label{eq:I_LF}
\bar{I}_{\nu}(z) = \int_{L_{\rm min}}^{L_{\rm max}} dL \,\Phi(L) \frac{L}{4\pi D_{L}^2}\,y(z)D_{A}^2.
\ee
Here we take $L_{\rm min}=10^6$ $L_{\sun}$ and $L_{\rm max}=10^{12}$ $L_{\sun}$, and $D_L$ and $D_A$ are the luminosity and comoving angular diameter distance, respectively. The factor $y(z)=dr/d\nu=\lambda_{\rm line}(1+z)^2/H(z)$, where $r$ is the comoving distance, $\lambda_{\rm line}$ is the rest-frame wavelength of emission line, and $H(z)$ is the Hubble parameter at $z$. The uncertainty of $\bar{I}_{\nu}$ is evaluated from the uncertainties in the observed LFs for each line, which is based on the errors of the LF fitting parameters. 

\subsection{Mean intensity from the SFR}

\begin{table*}[!t]
\caption{The fitting parameters for the SFR(M) derived from simulations in Guo et al. (2013). }
\vspace{-3mm}
\begin{center}
\begin{tabular}{  l   c   c   c   c   c   c   c   c}
\hline\hline
           & $z=1$ & $z=1.4$ & $z=1.8$ & $z=2.2$ & $z=2.7$ & $z=3.3$ & $z=4.0$ & $z=4.8$ \\
\hline \vspace{-2mm} \\
$a$ & -7.90 & -7.70 & -7.50 & -7.10 & -6.78 & -6.30 & -6.15 & -5.90 \\ 
$b$ & 2.50 & 2.49 & 2.49 & 2.42 & 2.36 & 2.25 & 2.25 & 2.25 \\
$c$ & -2.18 & -2.18 & -2.25 & -2.10 & -2.20 & -2.20 & -2.20 & -2.20  \\
\vspace{-2mm} \\
\hline
\end{tabular}
\end{center}
\vspace{-4mm}
\begin{center}
$-$ $M_1$ and $M_2$ in Eq.(\ref{eq:SFR_sim}) are fixed to be $1.0\times10^8$ and $4.0\times10^{11}$ $M_{\sun}$, respectively, for all redshifts.\\
\end{center}
\label{tab:SFR_sim}
\end{table*}

The strengths of the H$\alpha$\,6563$\rm\AA$, [OIII]\,$5007\rm\AA$, [OII]\,$3727\rm\AA$ and H$\beta$\,$4861\rm\AA$ lines are tightly related to the SFR of galaxies, which provides another way to evaluate their mean intensities through the $L_{\rm line}-{\rm SFR}$ relation. The relations are given by \citep{Kennicutt98, Ly07, Gong14}

\ba
{\rm SFR}\,(M_{\sun}{\rm yr^{-1}}) &=& (7.9\pm2.4)\times 10^{-42} L_{\rm H\alpha},\label{eq:SFR_Ha}\\
{\rm SFR}\,(M_{\sun}{\rm yr^{-1}}) &=& (7.6\pm3.7)\times 10^{-42} L_{\rm [OIII]},\label{eq:SFR_OIII}\\
{\rm SFR}\,(M_{\sun}{\rm yr^{-1}}) &=& (1.4\pm 0.4)\times 10^{-41} L_{\rm [OII]}.\label{eq:SFR_OII}
\ea
For the H$\beta$ line, we take the line ratio H$\beta/$H$\alpha=0.35$ \citep{Osterbrock06}, which is found to be in good agreement with observations and simulations, as we discuss later. 

\begin{figure}[t]
\includegraphics[scale = 0.44]{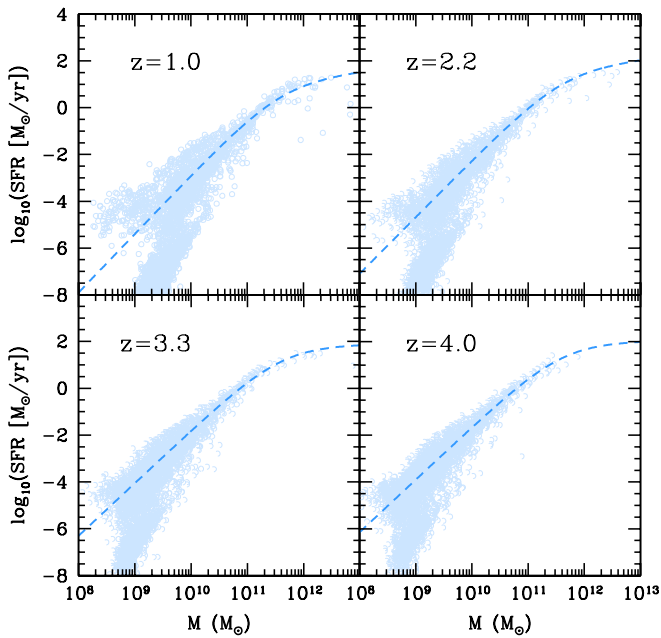}
\caption{\label{fig:SFR_sim} The SFR derived from simulations in \cite{Guo13}. The SFR data are in light blue circles and the best fitting results are in blue dashed curves. We can find the uncertainties of the SFR increase for low-mass halos with $M<10^{11}$ $M_{\sun}$, and the star formation is quenched at $M<10^8$ $M_{\sun}$.}
\end{figure}

\begin{figure*}[t]
\epsscale{1.9}
\centerline{
\resizebox{!}{!}{\includegraphics[scale=0.42]{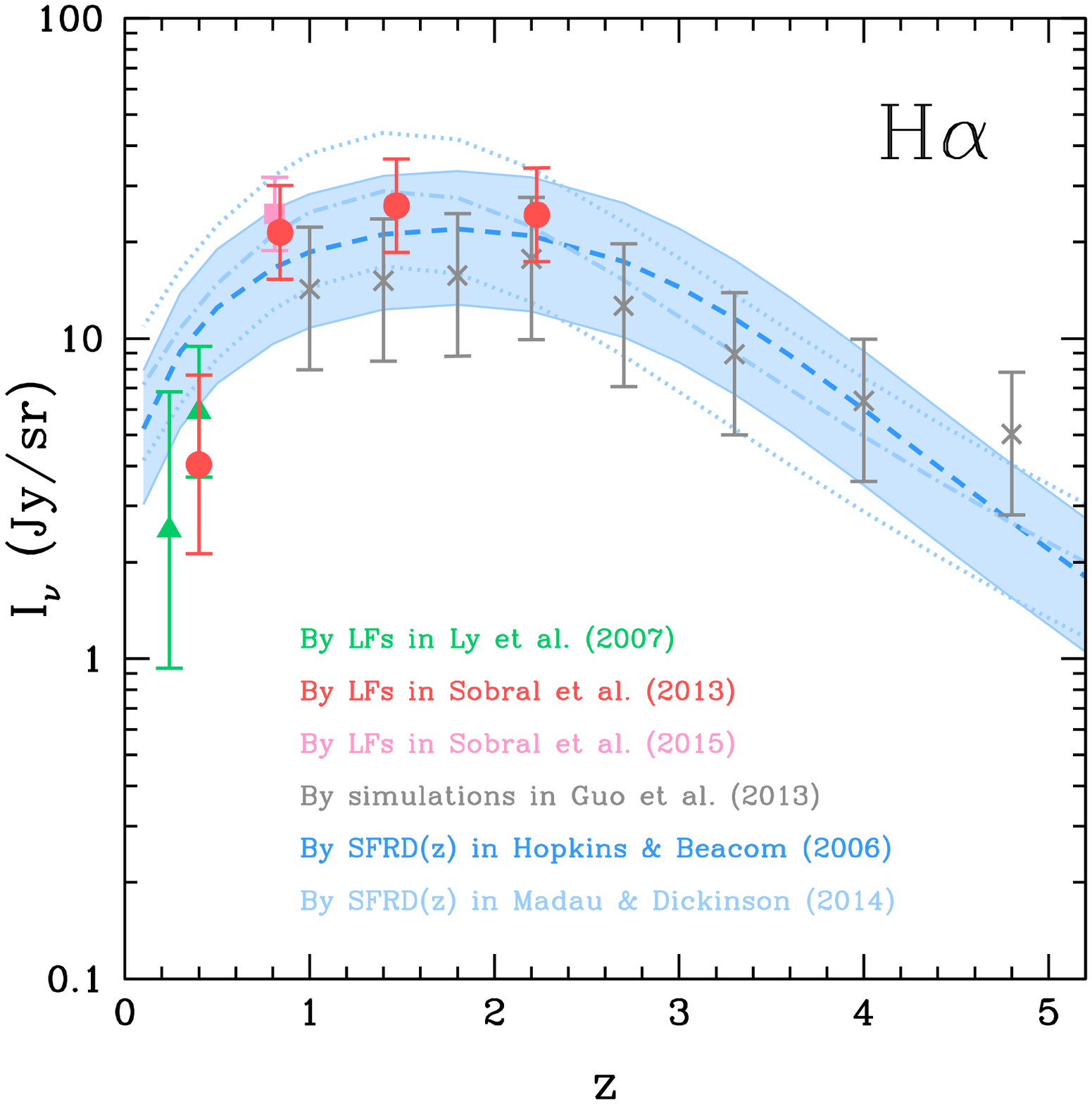}}
\resizebox{!}{!}{\includegraphics[scale=0.42]{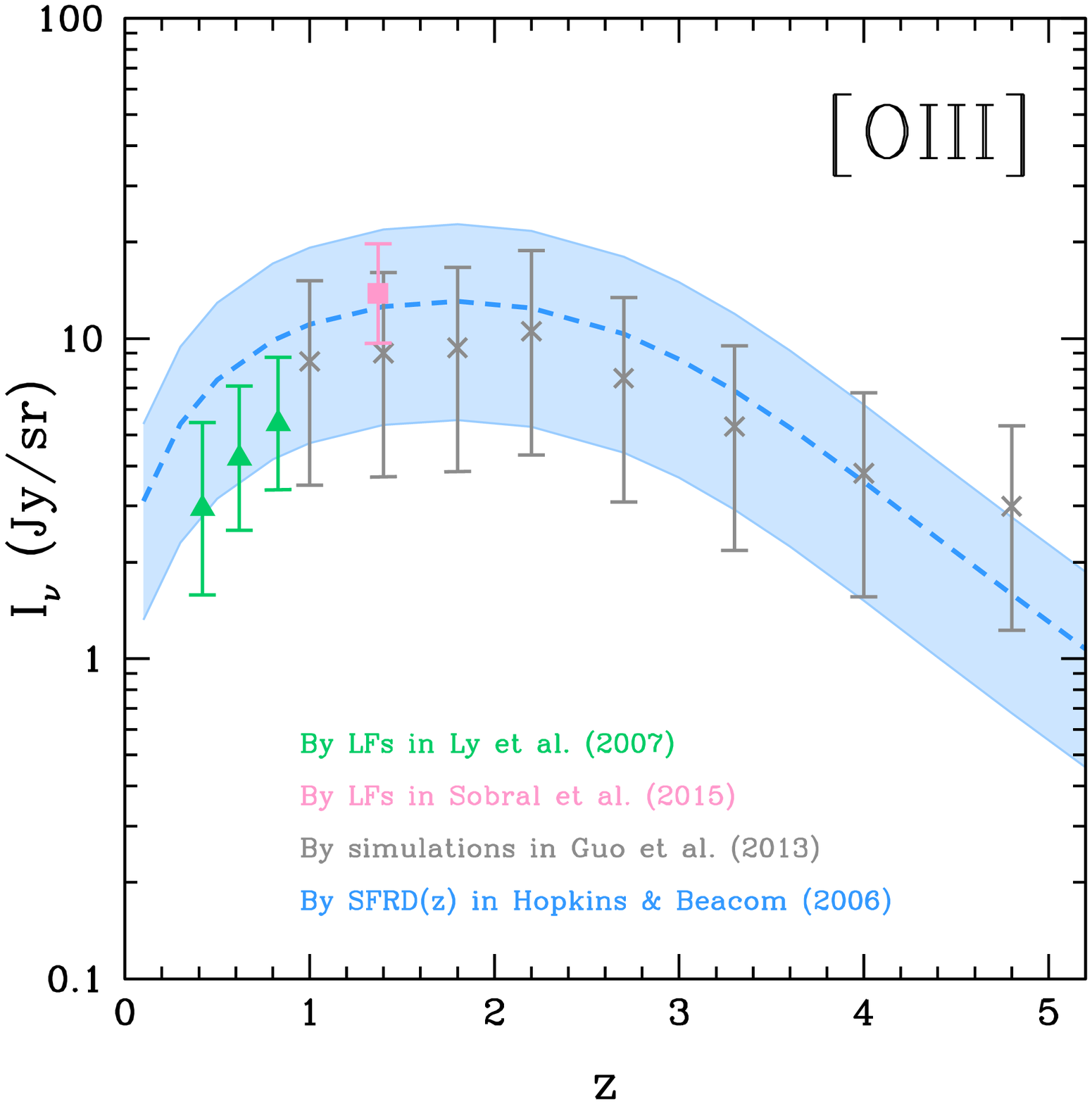}}
}
\centerline{
\resizebox{!}{!}{\includegraphics[scale=0.42]{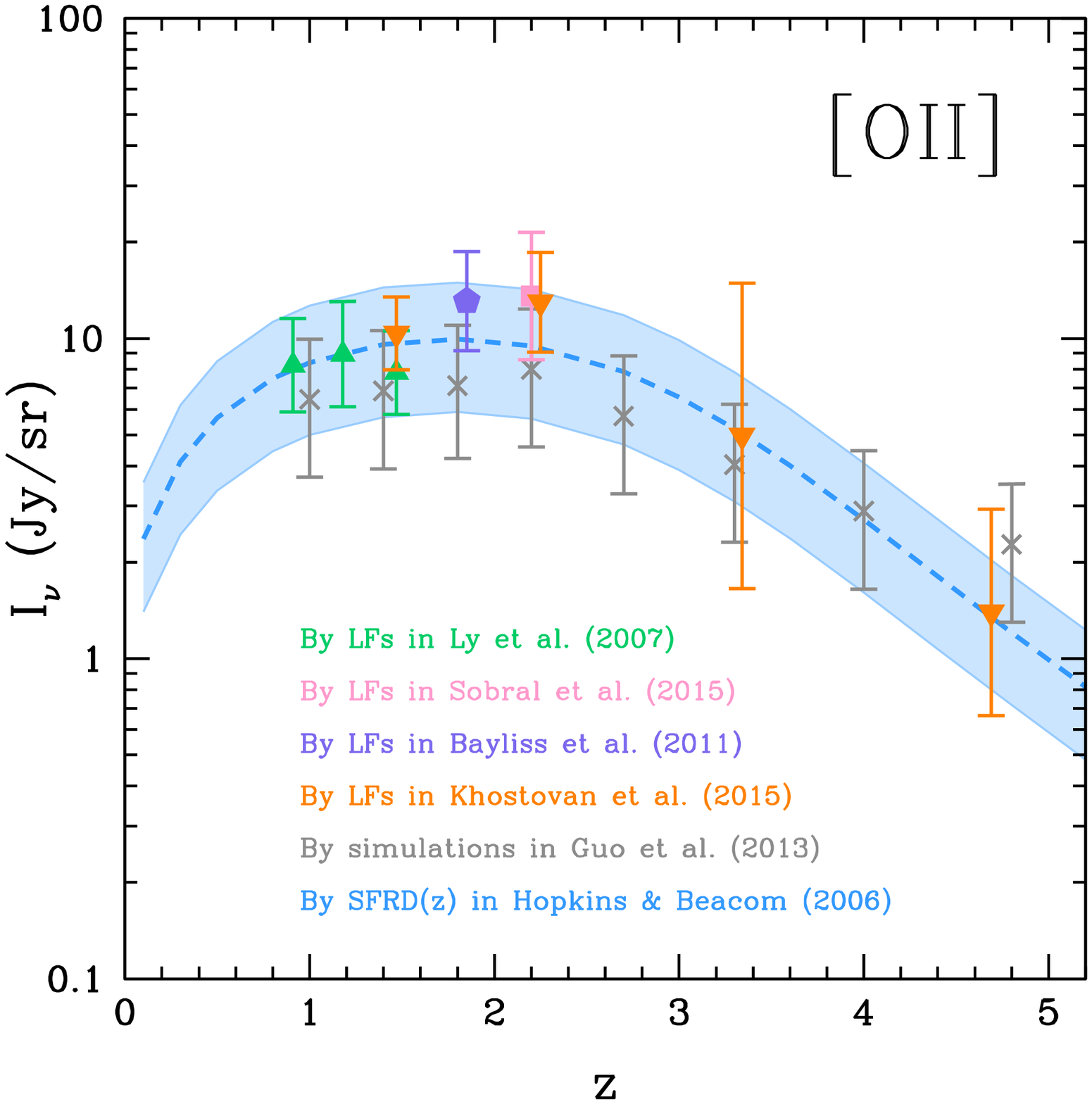}}
\resizebox{!}{!}{\includegraphics[scale=0.42]{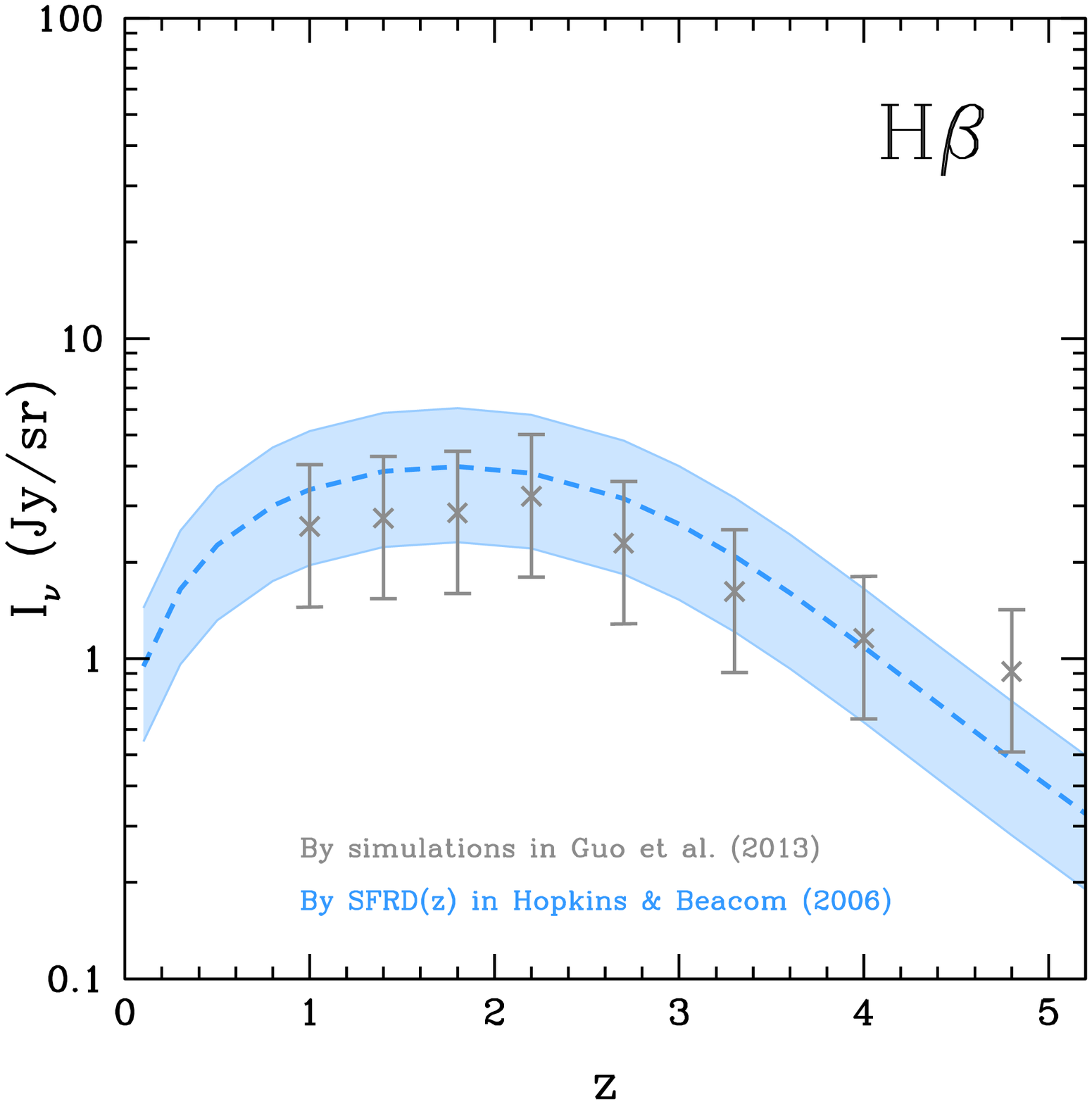}}
}
\epsscale{1.0}
\caption{\label{fig:Inus} Mean intensities of H$\alpha$, [OIII], [OII] and H$\beta$ lines at $z<5$, with no correction for dust extinction applied. The colored points are derived from observed LFs \citep{Ly07,Bayliss11,Sobral13,Sobral15,Khostovan15}. The gray crosses show the results from the simulations in \cite{Guo13}. The blue dashed curves with blue bands give the results of the SFRD in \cite{Hopkins06} for mean intensities and uncertainties. For comparison, the light blue dash-dotted and dotted curves denote the results with errors from the SFRD given by \cite{Madau14}. We find the results from the three methods (i.e. LFs, SFR simulations and observed SFRD) are basically consistent with one another in 1$\sigma$ C.L.}
\end{figure*}

\begin{figure}[t]
\includegraphics[scale = 0.44]{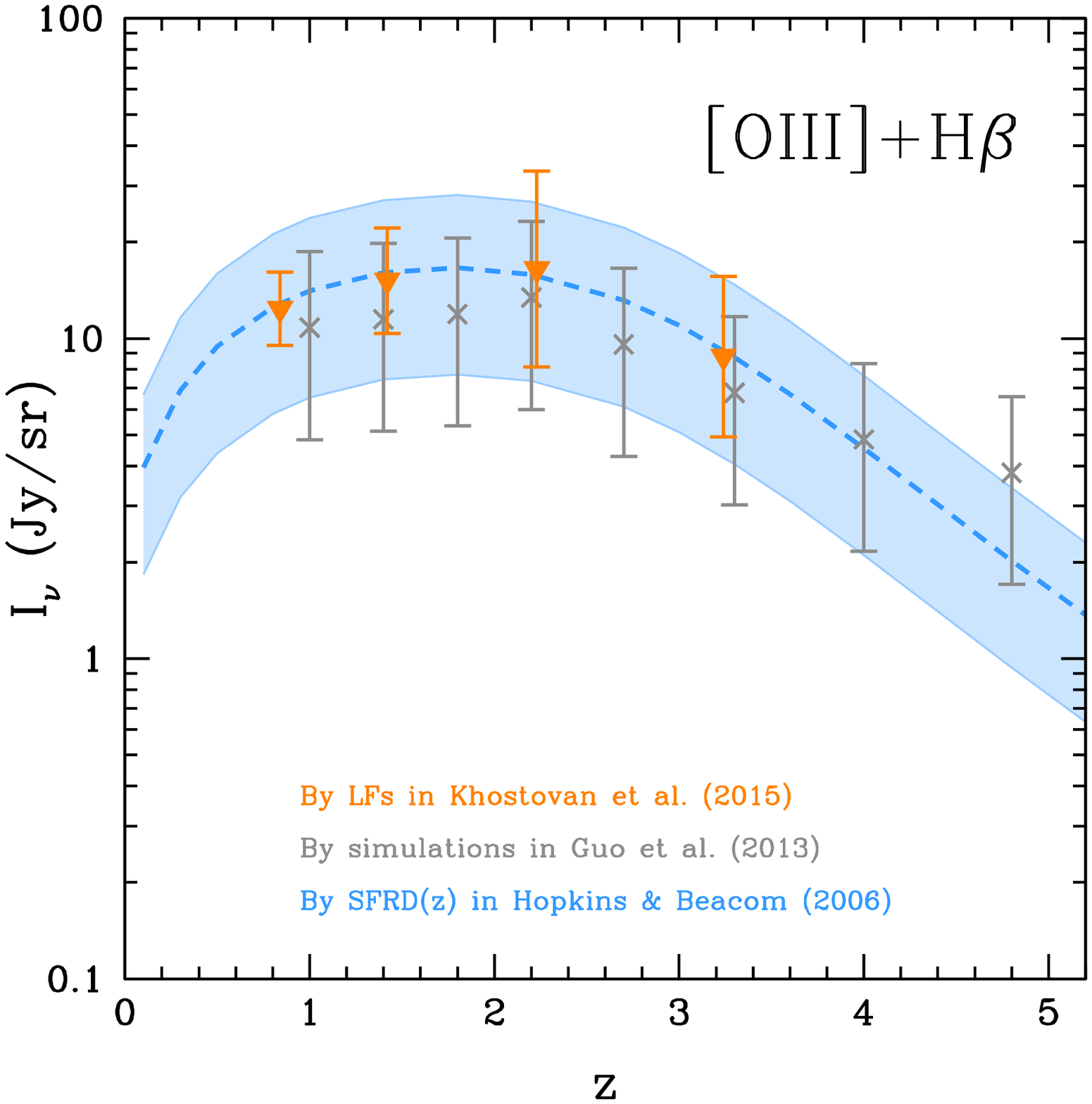}
\caption{\label{fig:Inu_OIII_Hb} Mean intensity of [OIII]+H$\beta$ without corrections for dust extinction. We find the results are consistent with one another, indicating the relatively simple H$\beta$ intensity estimate is reliable.}
\end{figure}

The SFR can be evaluated by two means, both of which are associated to the halo mass $M$. First, we derive the SFR$(M)$ from the simulations in \cite{Guo13}. These Millennium and Millennium II simulations are based on the first-year WMAP results for a $\Lambda$CDM universe with cosmological parameters rescaled to be consistent with the seven-year WMAP data \citep{Komatsu11}. A semi-analytic galaxy and star formation model is employed in the simulations, which is calibrated by the SDSS data \citep{Guo13}. We use the following formula to fit the SFR($M$) at different redshifts
\be \label{eq:SFR_sim}
{\rm SFR}\,(M) = 10^a \left( \frac{M}{M_1}\right)^b\left( 1+\frac{M}{M_2}\right)^c,
\ee
where $a$, $b$, $c$, $M_1$ and $M_2$ are free parameters. In Figure \ref{fig:SFR_sim}, as examples, we show the SFR($M$) data (light blue circles) derived from the simulations and the best fitting results (blue dashed curves) at $z=1.0$, 2.2, 3.3 and 4.0. The fitting values of these parameters at $z=1.0$, 1.4, 1.8, 2.2, 2.7, 3.3, 4.0 and 4.8 are listed in Table \ref{tab:SFR_sim}. We find the SFR becomes flat at $M\gtrsim10^{12}$ $M_{\sun}$ for $z<5$. The uncertainty of SFR-$M$ increases for low-mass halos and there is almost no star formation for $M<10^8$ $M_{\sun}$. 

Besides, we also derive the SFR(M) from observational results of the cosmic star formation rate density (SFRD). Following \cite{Hopkins06}, we take the fitting formula given by \citep{Cole01}
\be
{\rm SFRD}(z) = \frac{a+bz}{1+(z/c)^d}\ (\,M_{\sun}{\rm yr^{-1} Mpc^{-3}}h),
\ee
where $a=0.0118$, $b=0.08$, $c=3.3$ and $d=5.2$ for the initial mass function proposed in \cite{Baldry03}. We also assess the SFRD(z) suggested by \cite{Madau14}, and we find that the results derived from the two SFRDs are similar.  As a theoretical discussion, we decide to take the SFRD given by \cite{Hopkins06}, since it seems more consistent with the mean intensities derived from both of the observed LFs and the simulations as we discuss later. For simplicity, we assume the SFR is proportional to halo mass $M$, which is a good approximation for $M<10^{12}$ $M_{\sun}$ as shown in Figure \ref{fig:SFR_sim}. The SFR is given by
\be
{\rm SFR}(M,z) = f_*(z)\frac{\Omega_{b}}{\Omega_{M}}\frac{1}{t_{s}}M,
\ee
where $t_s=10^8$ yr is the typical star formation timescale, and $f_*(z)$ is the star formation efficiency which can be determined by the relation of ${\rm SFRD}(z) = \int dM(dn/dM){\rm SFR}(M,z)$, where $dn/dM$ is the halo mass function \citep{Cooray02}.

With the SFR$(M,z)$ estimated by simulations and observed SFRD$(z)$, the luminosities $L(M,z)$ of H$\alpha$, [OIII], [OII] and H$\beta$ can be obtained by Equations (\ref{eq:SFR_Ha}), (\ref{eq:SFR_OIII}) and (\ref{eq:SFR_OII}), repectively. Then the mean intensity can be calculated from
\be \label{eq:I_SFR}
\bar{I}_{\nu}(z) = \int_{M_{\rm min}}^{M_{\rm max}} dM\frac{dn}{dM}\frac{L_{\rm line}(M,z)}{4\pi D_{\rm L}^2} y(z)D_{\rm A}^2,
\ee
where $M_{\rm min}=10^8$  $M_{\sun}h^{-1}$ and $M_{\rm max}=10^{13}$ $M_{\sun}h^{-1}$ are the minimum and maximum halo masses we assume. 

We note that this mean intensity is the intrinsic intensity without dust extinction, since it is directly derived from the SFR obtained by simulations and the SFRD$(z)$ corrected for extinction. To account for the effect of dust extinction, we assume the mean dust extinctions, which are averaged over $M_B$ magnitude, as $A_{\rm H\alpha}=1.0$ mag, $A_{\rm [OIII]}=1.32$ mag, $A_{\rm [OII]}=0.62$ mag, and $A_{\rm H\beta}=1.38$ mag \citep{Kennicutt98,Calzetti00,Hayashi13,Khostovan15}. Note that the values of $A_{\rm [OIII]}$ and $A_{\rm H\beta}$ are based on standard $A_{\rm H\alpha}=1.0$ mag, while $A_{\rm [OII]}$ is for $A_{\rm H\alpha}=0.35$ mag, obtained by comparing with the SFRD results \citep{Khostovan15}. When we estimate the uncertainty of the mean intensity $\bar{I}_{\nu}$, we consider both uncertainties of the relation of $L_{\rm line}$-SFR and the SFRD from observations for different redshifts.

\subsection{Results of mean intensity}

In Figure \ref{fig:Inus}, we show the estimated mean intensities of H$\alpha$, [OIII], [OII] and H$\beta$ lines at $z<5$. We find the mean intensities derived from the three methods (i.e. LFs, SFR simulations and observed SFRD) are mainly consistent in 1$\sigma$ confidence level (C.L.), except that the results for H$\alpha$ LFs at $z<0.5$ are somewhat lower and the results from the simulations around $z=4.8$ are slightly higher than the other two methods. As expected, the mean intensities follow the profile of SFRD$(z)$ with a peak around $z=2$. The intensity of H$\alpha$ line is stronger than the other lines at the same redshift, and vary from a few Jy/sr to $\sim$20-30 Jy/sr over the redshift range. The intensity of [OIII] line is comparable with H$\alpha$ and has larger uncertainty. The intensity of the [OII] line is lower than [OIII], and we find that the [OII] intensity from the SFRD is in good agreement with the LFs results over $1<z<5$. The H$\beta$ intensity is the smallest among the four lines, and is almost one order of magnitude lower than that of the H$\alpha$ line.

For the H$\beta$ line, good measurements of H$\beta$ LFs are not available for comparison, so we simply use H$\beta/$H$\alpha=0.35$ to derive the H$\beta$ intensity. In order to validate the results, we calculate the intensity of [OIII]+H$\beta$ and compare it with the results from observed [OIII]+H$\beta$ LFs in \cite{Khostovan15}. The result is shown in Figure \ref{fig:Inu_OIII_Hb}. We find that the [OIII]+H$\beta$ intensity obtained from the SFRD matches the result of LFs very well, which indicates that the H$\beta$ intensity we estimate is reasonable. 

By comparing the intensity results from LFs, SFR simulations and observed SFRD, we find that the intensities obtained by the SFRD$(z)$ of \cite{Hopkins06} are in good agreements with the other methods and convenient for theoretical estimation. Hence, we take the intensity results from the SFRD in our following discussion.

\section{Line intensity power spectrum}

\begin{figure*}[t]
\epsscale{1.9}
\centerline{
\resizebox{!}{!}{\includegraphics[scale=0.42]{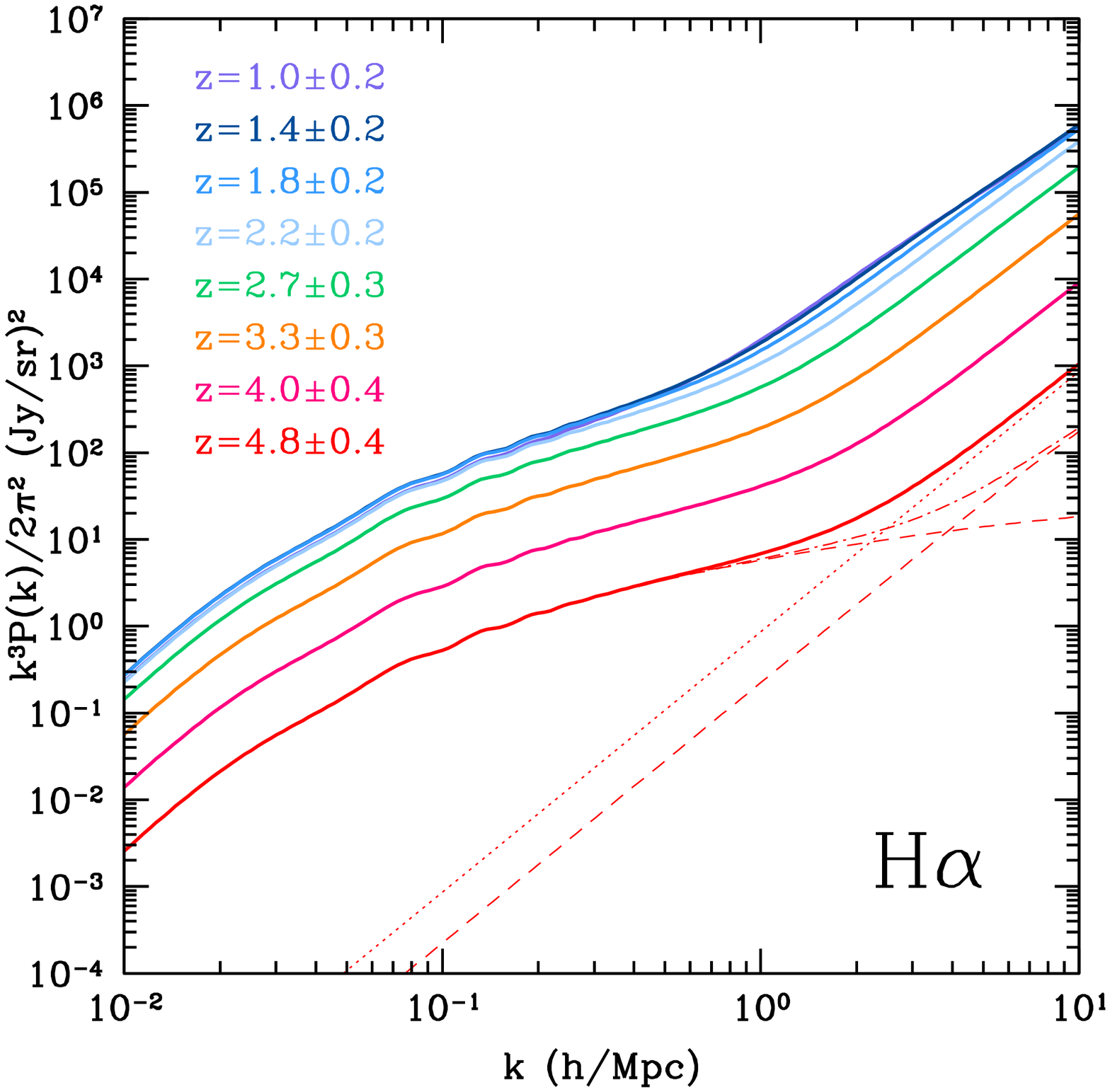}}
\resizebox{!}{!}{\includegraphics[scale=0.42]{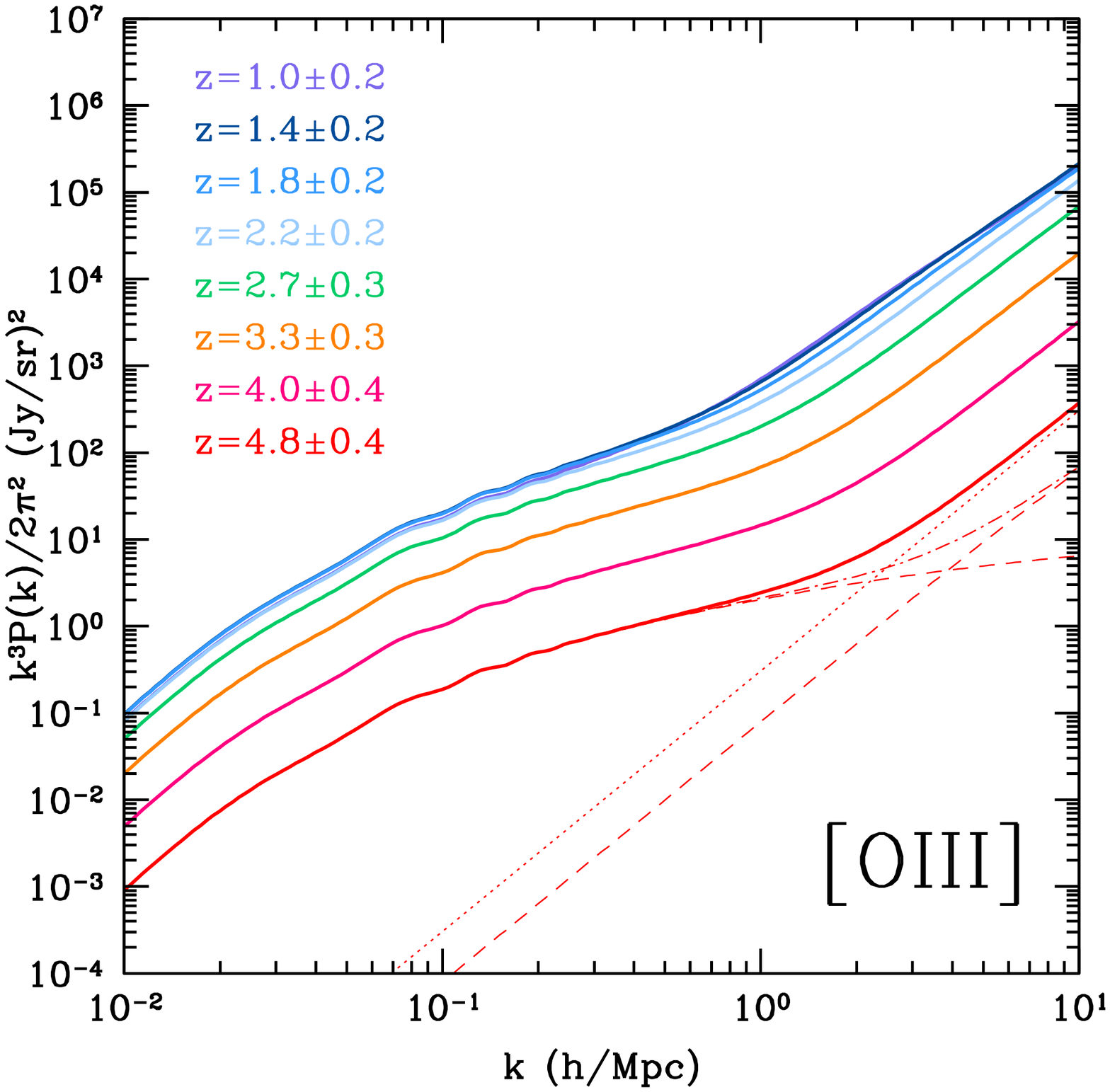}}
}
\centerline{
\resizebox{!}{!}{\includegraphics[scale=0.42]{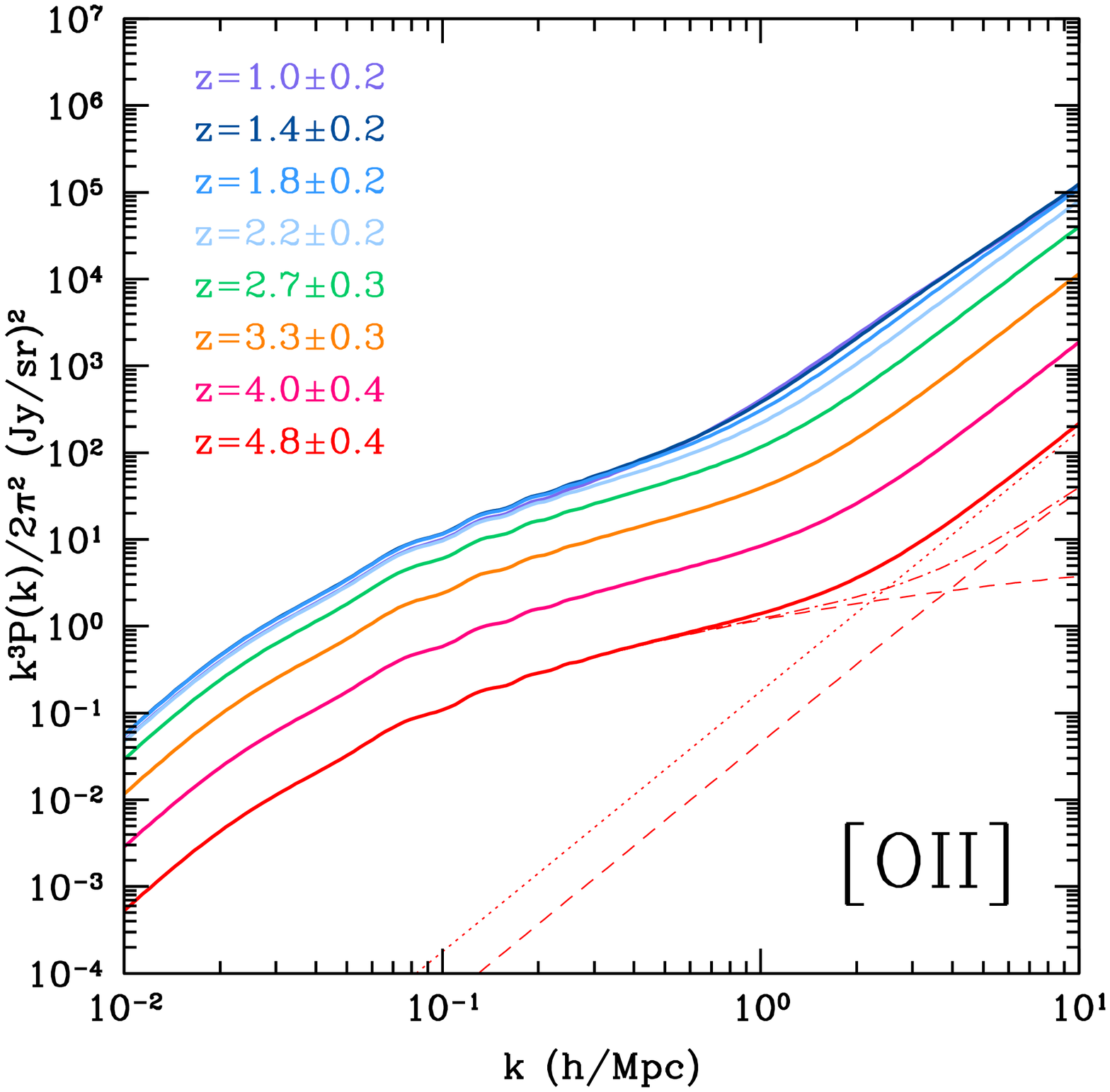}}
\resizebox{!}{!}{\includegraphics[scale=0.42]{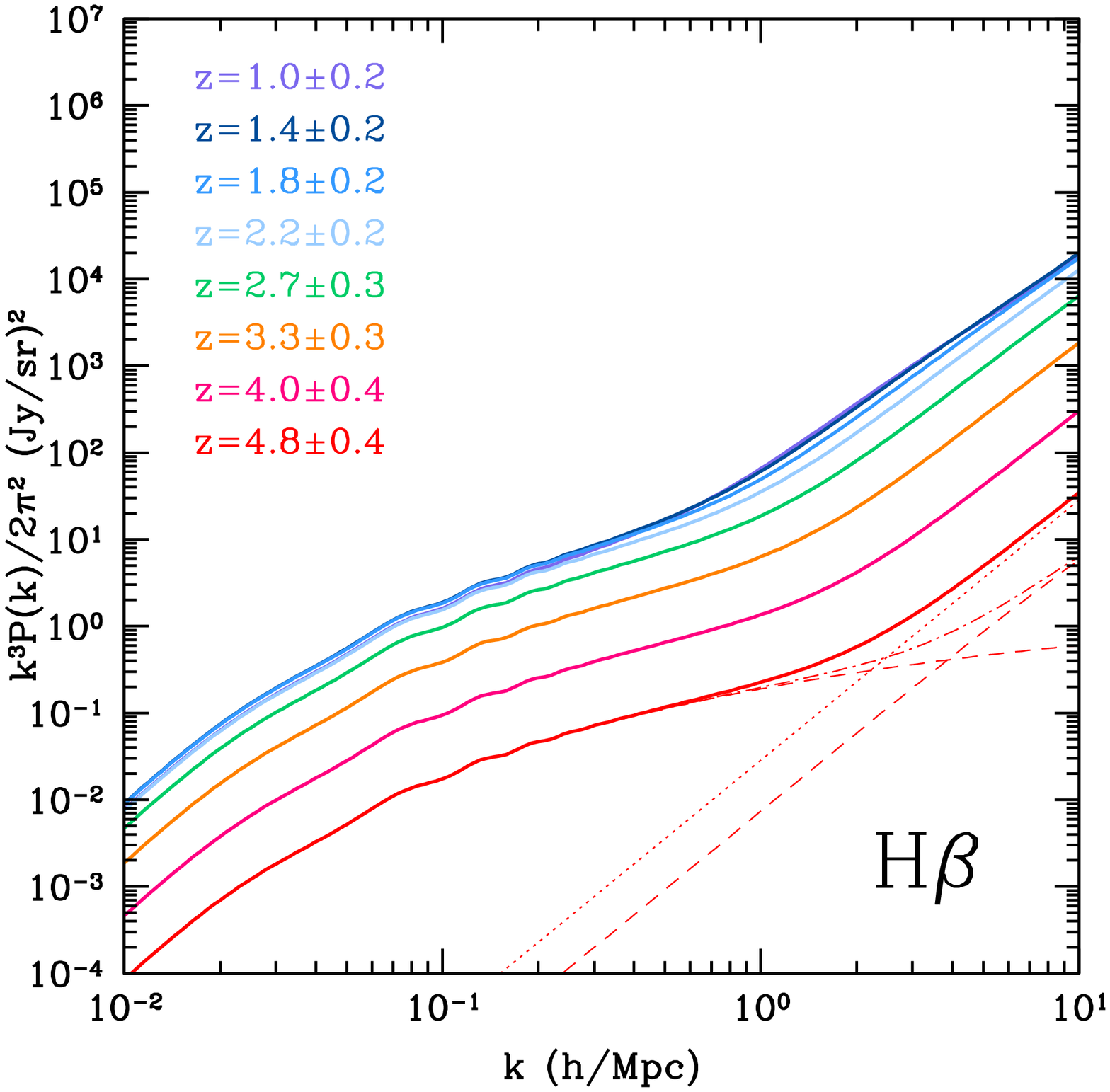}}
}
\epsscale{1.0}
\caption{\label{fig:PS} Intensity power spectra of H$\alpha$, [OIII], [OII] and H$\beta$ lines at different redshift ranges in $0.8\le z\le5.2$. These power spectra have included dust extinction effect. The dashed curves show the 1-halo and 2-halo terms for $z=4.8\pm0.4$, the dash-dotted lines are the clustering power spectra, and the dotted lines are the shot-noise power spectra.}
\end{figure*}

In this section, we estimate the 3-D intensity power spectra for H$\alpha$, [OIII], [OII] and H$\beta$ lines at different redshifts. Since the emission lines come from galaxies that trace the distribution of underlying dark matter, the fluctuation of line intensity can be expressed by
\be
\delta I_{\rm line} = \bar{b}_{\rm line}\,\bar{I}_{\rm line}\,\delta({\bf x}),
\ee
where $\bar{I}_{\rm line}$ is the mean intensity we estimate in the last section, $\delta(\bf x)$ is the over-density of dark matter field at position $\bf x$, and $\bar{b}_{\rm line}$ is the weighted galaxy bias by the luminosity of emission line, which is given by
\be \label{eq:b_line}
\bar{b}_{\rm line}(z)=\frac{\int^{M_{\rm max}}_{M_{\rm min}} dM \frac{dn}{dM}\, L_{\rm line}\, b(M,z)}{\int^{M_{\rm max}}_{M_{\rm min}} dM \frac{dn}{dM} \,L_{\rm line}},
\ee
where $b(M,z)$ is the halo bias \citep{Sheth99}. We can then calculate the clustering power spectrum of the line that traces galaxy clustering as
\be\label{eq:P_line}
P_{\rm line}^{\rm clus}(k,z) = \bar{b}_{\rm line}^{\,2} \,\bar{I}_{\rm line}^{\,2}\, P_{\delta \delta}(k,z).
\ee
Here $P_{\delta \delta}(k,z)$ is the matter power spectrum, and we use the halo model of to calculate it \citep{Cooray02}. For simplicity, we ignore redshift distortion in this work, and only focus on the isotropic spatial fluctuations. At small scales, shot noise is the dominant term, which is due to the discrete distribution of galaxies. The shot-noise power spectrum for an emission line is given by
\be\label{eq:Pshot_line}
P^{\rm shot}_{\rm line}(z) = \int_{M_{\rm min}}^{M_{\rm max}} dM \frac{dn}{dM} \left[\frac{L_{\rm line}}{4\pi D_{\rm L}^2}y(z)D_{\rm A}^2\right]^2.
\ee
For the result of using the LFs, the terms $dn/dM$ and $dM$ are replaced by $\Phi(L)$ and $dL$, respectively. The total power spectrum is $P_{\rm line}^{\rm tot}(k,z)=P_{\rm line}^{\rm clus}(k,z)+P^{\rm shot}_{\rm line}(z)$. 

In Figure \ref{fig:PS}, we show the power spectra of H$\alpha$, [OIII], [OII] and H$\beta$ lines at $z=1.0\pm0.2$, $1.4\pm0.2$, $1.8\pm0.2$, $2.2\pm0.2$, $2.7\pm0.3$, $3.3\pm0.3$, $4.0\pm0.4$ and $4.8\pm0.4$ including the effects of dust extinction. We assume there is no evolution for the power spectra in each redshift interval. These power spectra also can be seen as the average power spectra over the corresponding redshift intervals, since we find that they have similar amplitudes and shapes to the average ones. As expected, the power spectra of the H$\alpha$ line is higher than the other three lines at the same redshift, since its mean intensity is the largest. The power spectrum of [OIII] is larger than [OII], and H$\beta$ has the lowest power spectrum. We also find that, for all the four lines, the amplitudes of power spectra at $1\lesssim z\lesssim2.2$ are quite similar to one another, no matter the redshift. This is because that the mean intensity for each line peaks at $z\sim2$ and the bias in Equation (\ref{eq:b_line}) is larger at higher redshifts, which compensates for the decrease of the matter power spectrum at higher redshifts. However, at $z\gtrsim2.2$, the line power spectra decrease quickly as the redshift increases, since the mean intensity and the matter power spectrum both decrease. The clustering power spectrum is about two orders of magnitude higher at $1\lesssim z\lesssim2.2$ than that at $z\sim 5$, while it is almost three orders of magnitude higher for the shot-noise power spectra at the same redshifts. As we discuss in the next section, this leads to large differences in the detectability of the lines as a function of redshift.

\section{Detectability of the lines}

In this section, we study the detectability of the H$\alpha$, [OIII], [OII] and H$\beta$ lines. We first discuss contamination from foreground  lines at lower redshifts and the method of removal, and then explore the detectability of the four lines with the proposed SPHEREx experiment.

\subsection{Contamination of foreground emission lines}

\begin{figure*}[t]
\epsscale{1.9}
\centerline{
\resizebox{!}{!}{\includegraphics[scale=0.42]{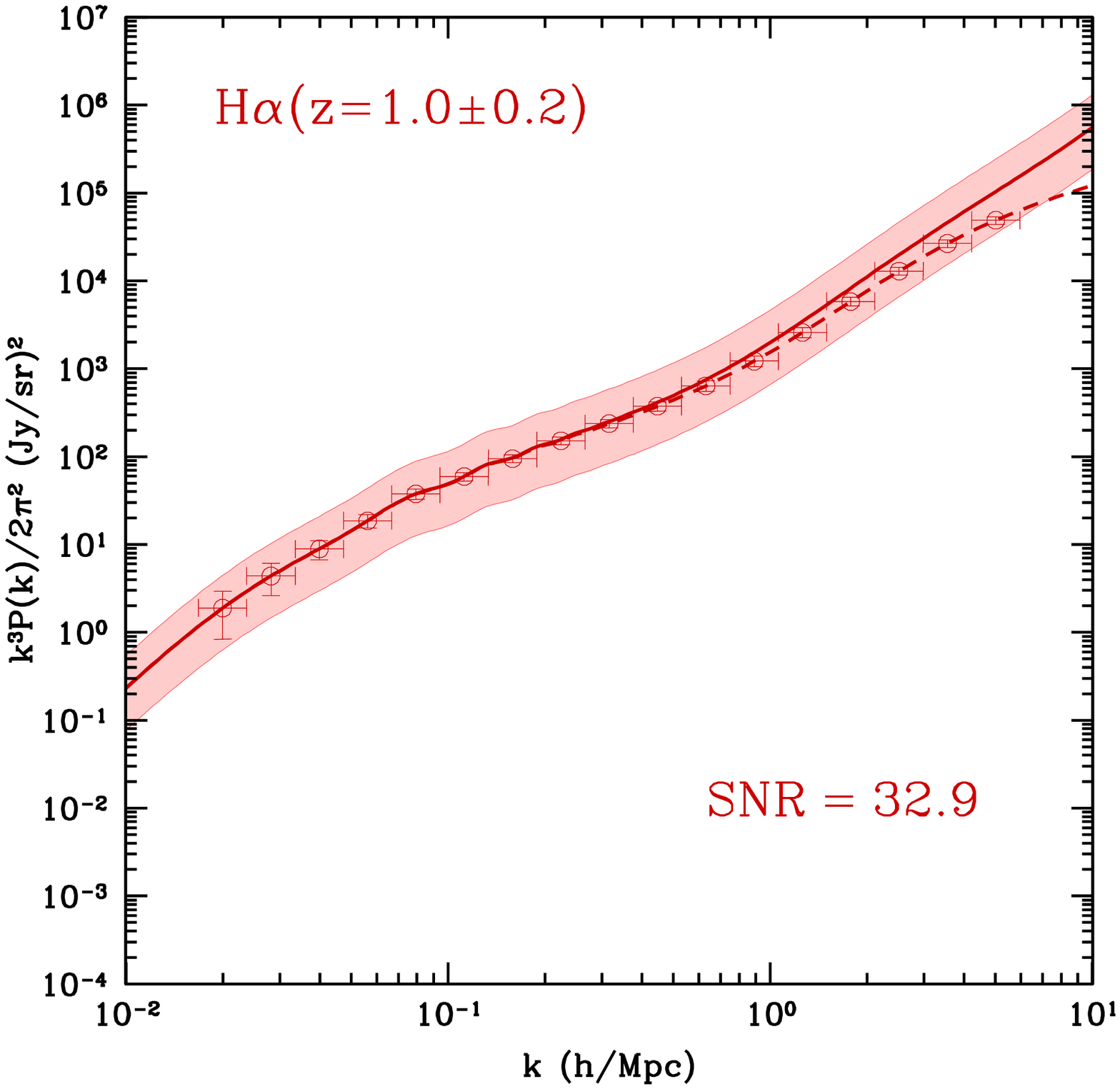}}
\resizebox{!}{!}{\includegraphics[scale=0.42]{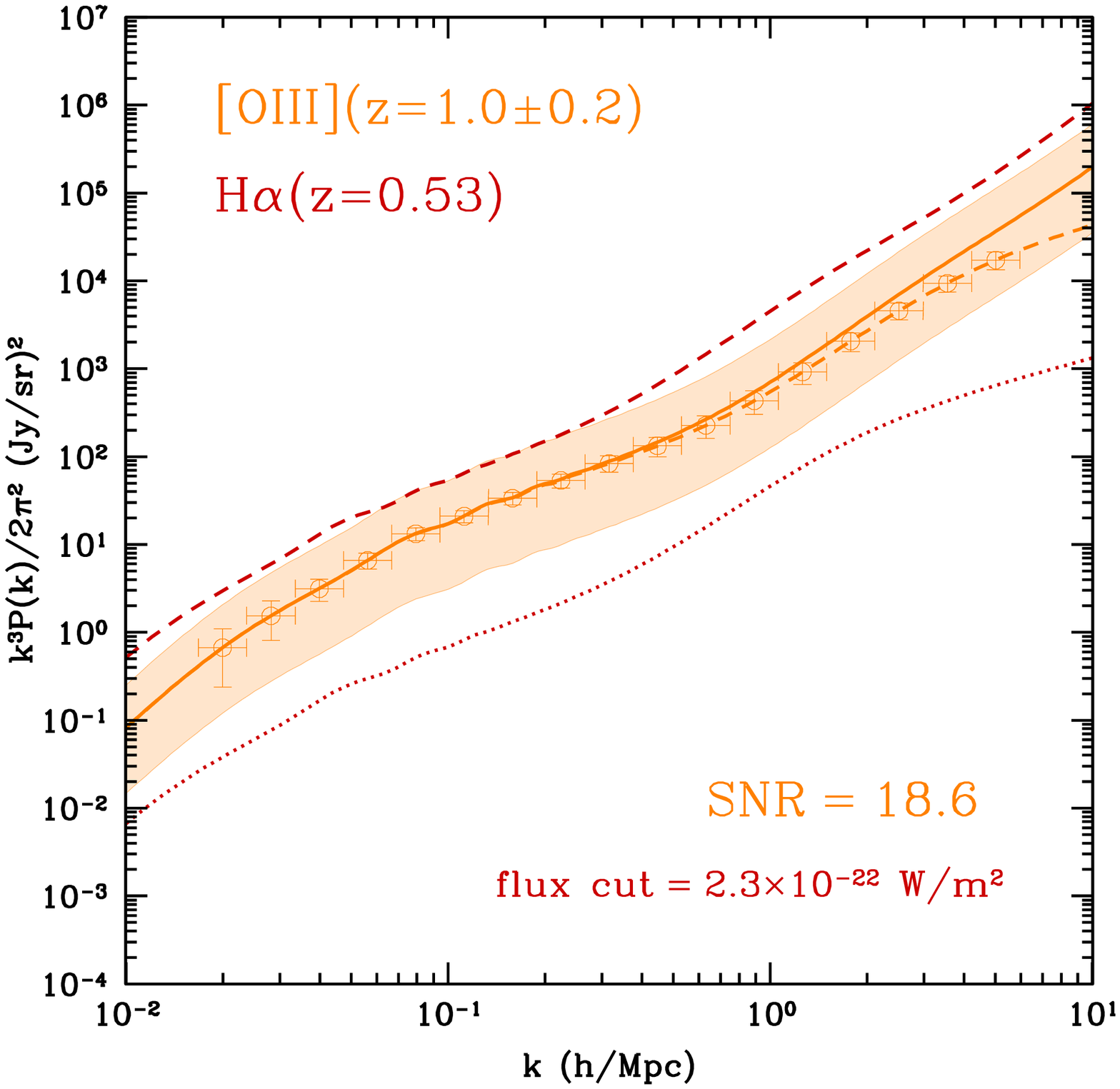}}
}
\centerline{
\resizebox{!}{!}{\includegraphics[scale=0.42]{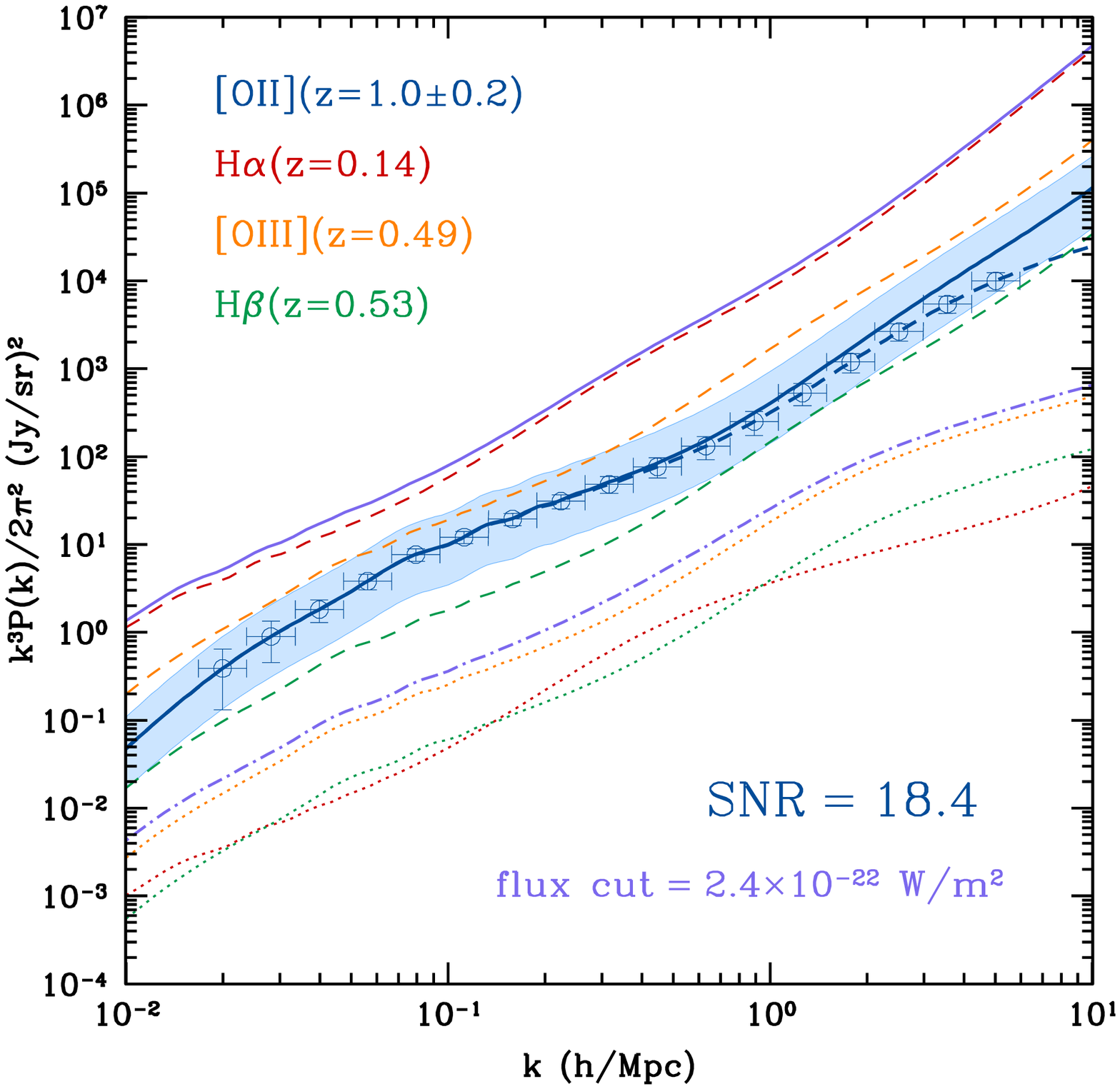}}
\resizebox{!}{!}{\includegraphics[scale=0.42]{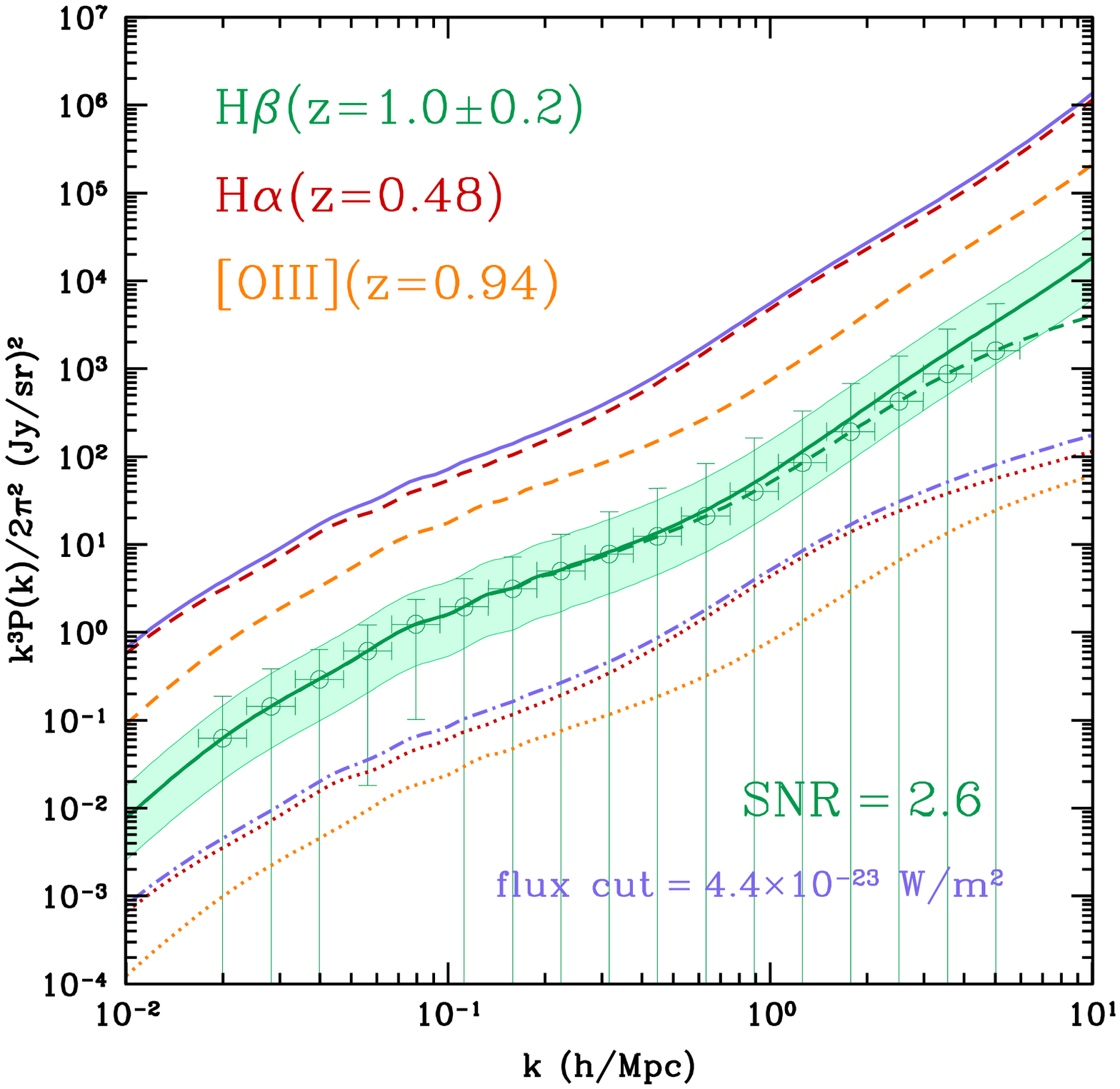}}
}
\epsscale{1.0}
\caption{\label{fig:PS_com} Intensity power spectra of H$\alpha$, [OIII], [OII] and H$\beta$ lines at $z=1.0\pm0.2$ with the foreground contaminants. $Upper$ $left$: The H$\alpha$ power spectrum is shown; There are no bright foreground lines that can affect the H$\alpha$ signal. The solid and dashed curves denote the total and clustering power spectra, respectively. $Upper$ $right$: The [OIII] line contaminated by the low-$z$ H$\alpha$ line. The red dashed line is the projected foreground $P^p_{{\rm H}\alpha}$, and red dotted line denotes the $P^p_{{\rm H}\alpha}$ after a flux cut at $2.3\times10^{-22}$ $\rm W/m^2$. $Lower$ $left$: The [OII] line contaminated by foreground H$\alpha$, [OIII] and H$\beta$ lines. The purple solid and dash-dotted curves are the sum of the foreground lines before and after a flux cut at $2.4\times10^{-22}$ $\rm W/m^2$. $Lower$ $right$: The H$\beta$ line contaminated by low-$z$ H$\alpha$ and [OIII] lines. We also show the error bars and SNR of the clustering power spectrum for each signal line for the SPHEREx experiment.}
\end{figure*}

\begin{figure*}[t]
\epsscale{1.9}
\centerline{
\resizebox{!}{!}{\includegraphics[scale=0.42]{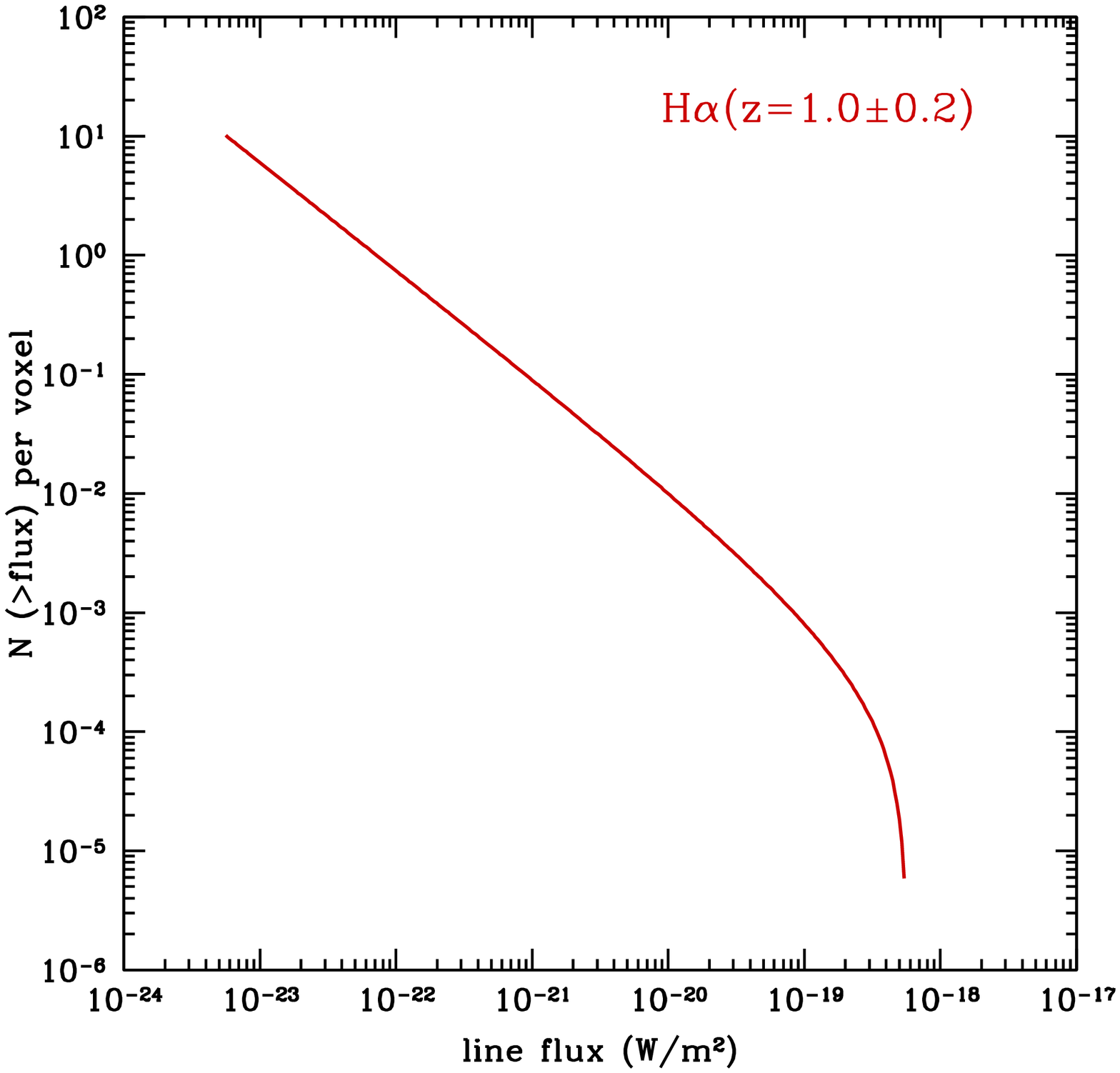}}
\resizebox{!}{!}{\includegraphics[scale=0.42]{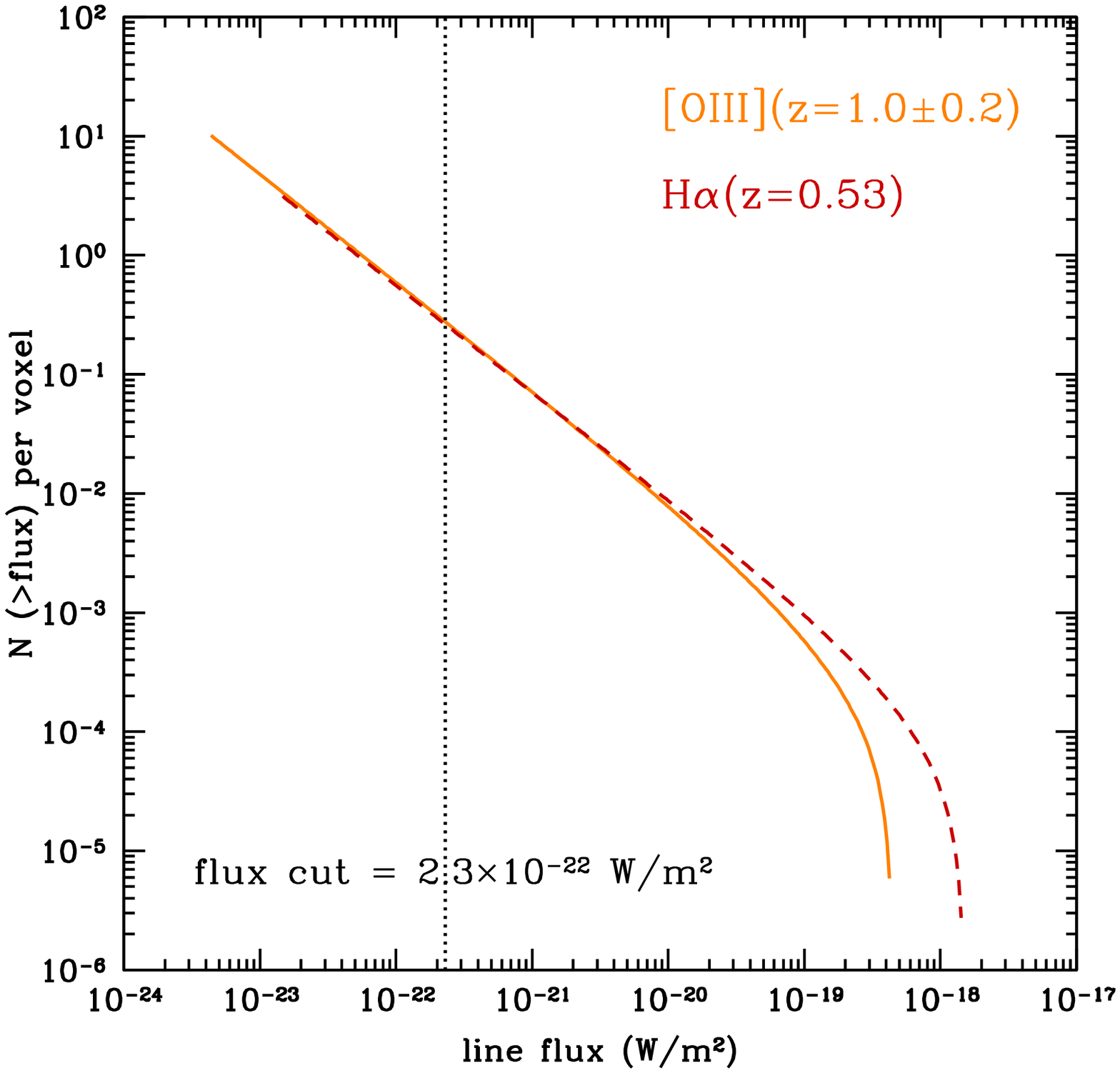}}
}
\centerline{
\resizebox{!}{!}{\includegraphics[scale=0.42]{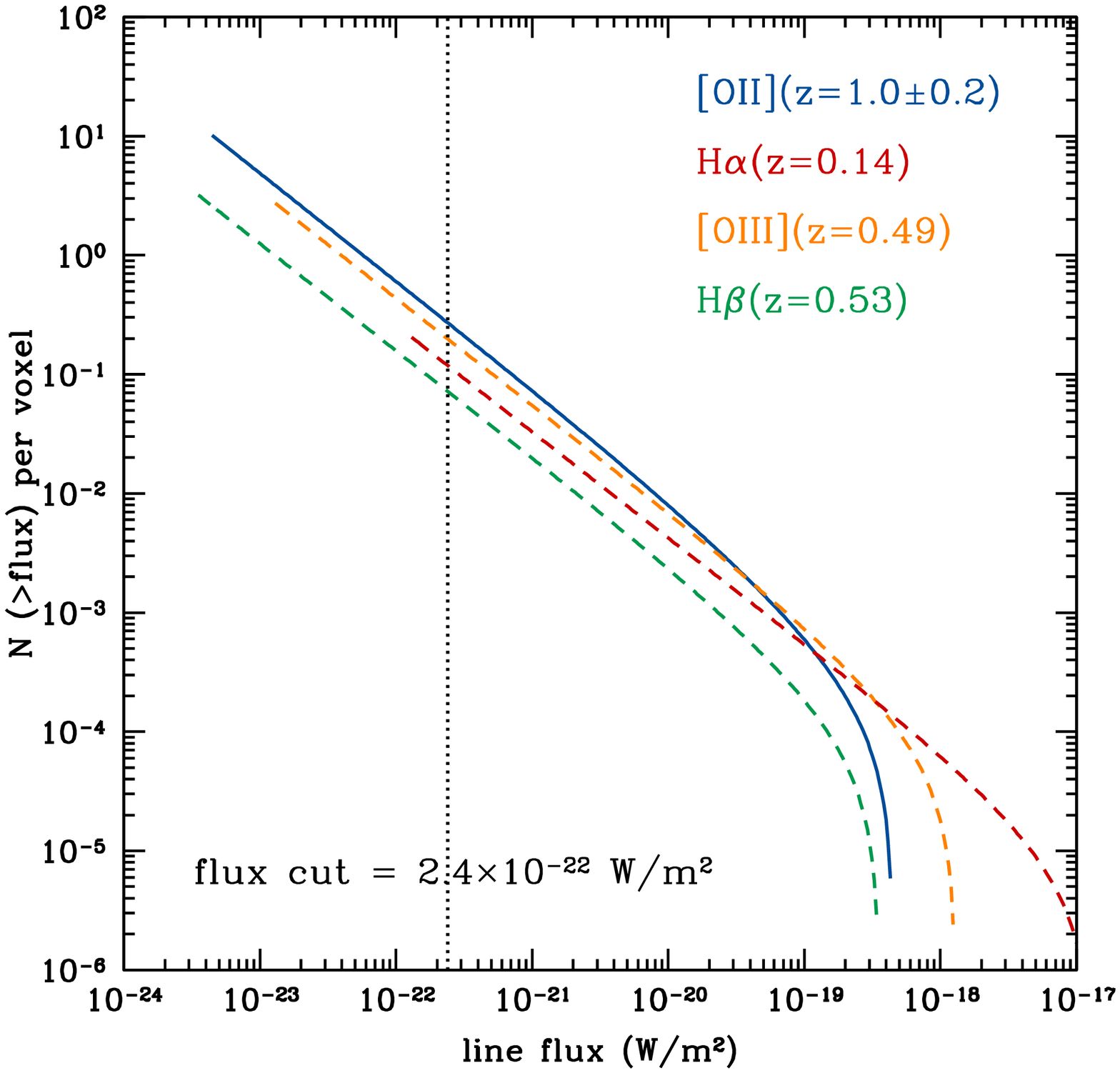}}
\resizebox{!}{!}{\includegraphics[scale=0.42]{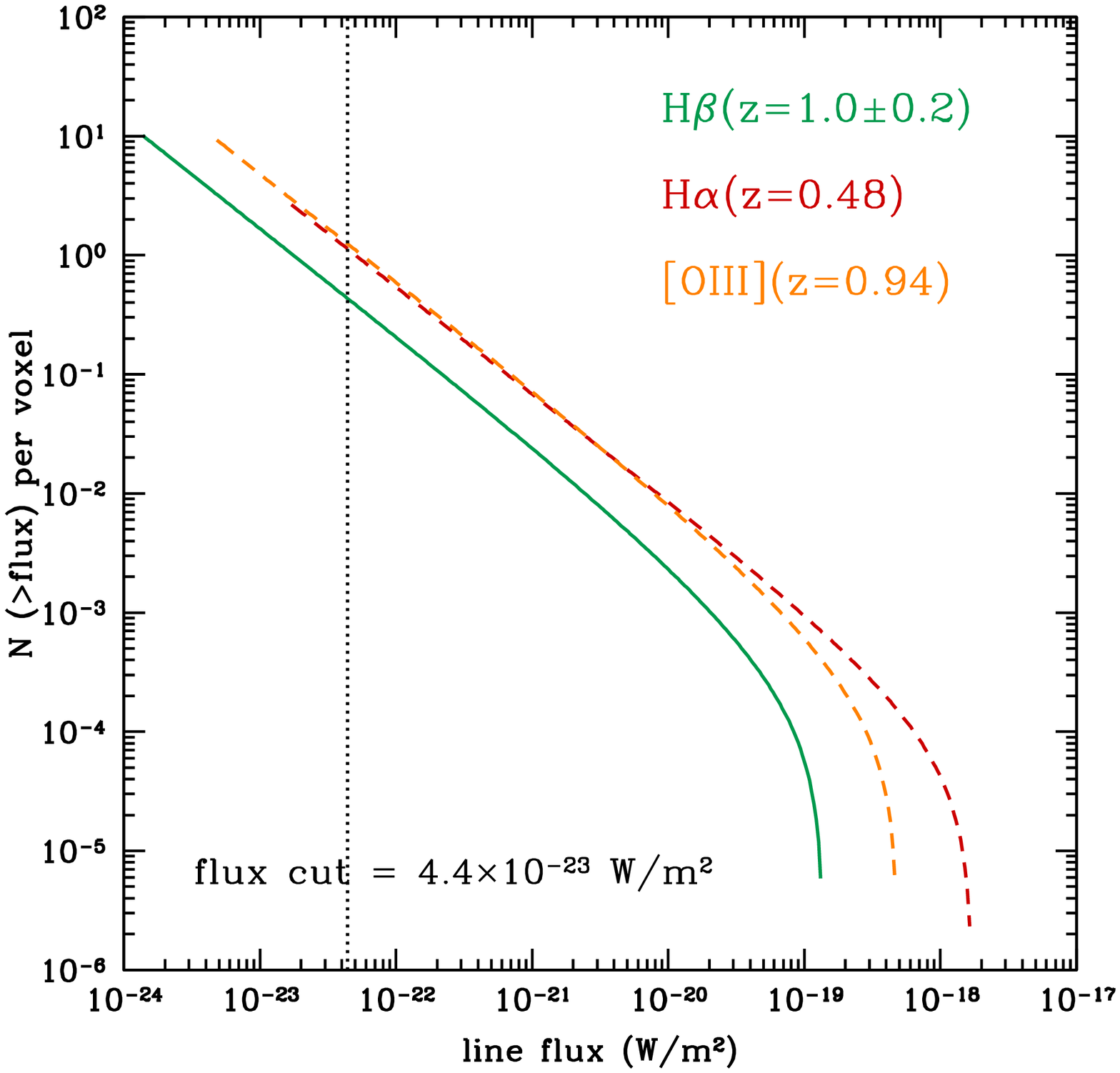}}
}
\epsscale{1.0}\
\caption{\label{fig:N_f} Total number of sources in a SPHEREx survey voxel whose flux is greater than a given value. The solid and dashed curves denote the number of signal and foreground lines, respectively. The flux cuts listed in Table \ref{tab:SNR_fc} are shown by vertical dotted lines.}
\end{figure*}

\begin{table*}[!t]
\caption{SNRs, flux cuts and percentages of removed pixels for H$\alpha$, [OIII], [OII] and H$\beta$ line intensity mapping at $0.8\le z\le5.2$.}
\vspace{-4mm}
\begin{center}
\begin{tabular}{  l  l  c  c   c   c   c   c   c   c   c}
\hline\hline
 Line        &  & $z\sim1.0$ & $z\sim1.4$ & $z\sim1.8$ & $z\sim2.2$ & $z\sim2.7$ & $z\sim3.3$ & $z\sim4.0$ & $z\sim4.8$ \\
\hline \vspace{-2mm}\\
                &SNR & 32.9 & 44.0 & 33.6 & 29.4 & 39.6 & 20.0 & 7.1 & 1.3 \\ 
H$\alpha$&flux cut ($\rm W/m^2$)& - & - & - & - & - & - & - & - \\
                &\% of removed pix.& - & - & - & - & - & - & - & -  \\
 \hline   
                &SNR & 18.6 & 13.0 & 19.7 & 14.4 & 11.5 & 4.8 & 2.6 & 0.5 \\              
{\rm [OIII]}&flux cut ($\rm W/m^2$) & $2.3\times10^{-22}$ & $1.4\times10^{-22}$ & $8.4\times10^{-23}$ & $6.0\times10^{-23}$ & $2.4\times10^{-23}$ & $6.8\times10^{-24}$ & $2.8\times10^{-24}$ & $1.3\times10^{-24}$ \\
                &\% of removed pix.& 3\% & 4\% & 5\% & 6\% & 10\% & 21\% & 42\% & 55\%  \\
 \hline   
                &SNR & 18.4 & 11.4 & 8.2 & 6.4 & 8.4 & 3.2 & 0.8 & 0.1 \\
{\rm [OII]}&flux cut ($\rm W/m^2$) & $2.4\times10^{-22}$ & $8.6\times10^{-23}$ & $4.0\times10^{-23}$ & $2.5\times10^{-23}$ & $9.9\times10^{-24}$ & $4.6\times10^{-24}$ & $2.4\times10^{-24}$ & $9.9\times10^{-25}$ \\
                &\% of removed pix.& 3\% & 5\% & 9\% & 20\% & 32\% & 43\% & 53\% & 67\%  \\
 \hline   
                &SNR & 2.6 & 1.8 & 2.9 & 2.1 & 1.4 & 0.5 & 0.3 & $<0.1$ \\ 
H$\beta$ &flux cut ($\rm W/m^2$) & $4.4\times10^{-23}$ & $2.6\times10^{-23}$ & $1.3\times10^{-23}$ & $9.2\times10^{-24}$ & $5.2\times10^{-24}$ & $3.0\times10^{-24}$ & $1.6\times10^{-24}$ & $8.8\times10^{-25}$ \\
                &\% of removed pix.& 12\% & 15\% & 21\% & 27\% & 38\% & 48\% & 65\% & 87\%  \\ \vspace{-2mm}\\
\hline
\end{tabular}
\end{center}
\vspace{-4mm}
\begin{center}
\end{center}
\label{tab:SNR_fc}
\end{table*}

In an intensity mapping survey, the detected pixel volume (or voxel) is characterized by the space and frequency resolutions of the instrument, whose directions are perpendicular and parallel to the line of sight, respectively. Hence, the different emission lines at different redshifts at the same observed frequency can be mixed together in a voxel. This effect is particularly important for measuring an emission line at high redshift, as other luminous lines at longer wavelengths at lower redshifts can contaminate the line at high redshift in which we are interested.

For the H$\alpha\,6563\rm\AA$ line, there is no emission line at longer wavelengths that can provide considerable contamination at lower redshifts, and so that it is a good tracer for studies of the SFRD and matter distribution in intensity mapping surveys. Here we ignore foreground contaminating lines for the H$\alpha$ line. For [OIII]$\,5007\rm\AA$, the main foreground line contaminant is H$\alpha$ line at lower redshifts. For the H$\beta\,4861\rm\AA$ line, we consider both low-$z$ H$\alpha$ and [OIII] lines as foreground contaminants. The [OII]$\,3727\rm\AA$ line has the shortest wavelength among the four luminous lines, so it can be contaminated by all the other three lines at lower redshifts. 

Following \cite{Gong14}, the observed 3-D power spectrum is the sum of the signal power spectrum and all the projected foreground line power spectra
\be
P_{\rm obs}(k,z) = P_s(k,z) + \sum_{i=1}^N P_f^{p,i}(k_f,z).
\ee
Here $k=\sqrt{k_{\perp}^2+k_{\parallel}^2}$ is the 3-D comoving wavenumber at the signal redshift $z$, where $k_{\perp}$ and $k_{\parallel}$ are the components which are perpendicular and parallel to the line of sight, respectively. The $k_f$ denotes the wavenumber corresponding to $k$ at the foreground redshift $z_f$, which is given by $k_f=\sqrt{A_{\perp}^2k_{\perp}^2+A_{\parallel}^2k_{\parallel}^2}$. $A_{\perp}$ and $A_{\parallel}$ are the factors to transfer $k$ to $k_f$, and we have $A_{\perp}=r_s/r_f$ and $A_{\parallel}=y_s/y_f$. The signal power spectrum $P_s(k,z)$ is the total power spectrum $P_{\rm line}^{\rm tot}(k,z)$ we derive in the last section. The projected foreground power spectrum $P^p_f(k_f,z)$ is the foreground power spectrum $P_f(k_f, z_f)$ projected to the signal redshift $z$, which takes the form
\be
P^p_f(k_f,z) = A_{\perp}^2A_{\parallel} P_f(k_f,z_f).
\ee 
The factor $A_{\perp}^2A_{\parallel}$ is caused by the expansion of volume elements when Fourier transforming the foreground correlation function at $z_f$ to the projected foreground power spectrum at $z$ \citep{Visbal10,Gong14}. We can find that the $P^p_f$ is not isotropic along and perpendicular to the direction of line of sight, even after ignoring the redshift distortion effect, since the values of factors $A_{\perp}$ and $A_{\parallel}$ are different given that $z_s\ne z_f$. This effect can help us to identify and remove the foregrounds in intensity mapping surveys \citep{Gong14,Lidz16}. In this work, we focus on the 3-D power spectrum, and we assume $k_1=k_2=k_{\parallel}$ where ${\bf k}=k_1\hat{{\bf i}}+k_2\hat{{\bf j}}+k_{\parallel}\hat{{\bf n}}$ and $k_{\perp}=\sqrt{k_1^2+k_2^2}$.

Since the foreground lines at lower redshifts are relatively brighter than the signal line at higher redshift, the most direct way to remove the foregrounds is to mask the bright pixels above some flux level. In Figure \ref{fig:PS_com}, we show the intensity power spectrum of H$\alpha$, [OIII], [OII] and H$\beta$ lines at $z=1.0\pm0.2$ and the foreground contaminating lines at lower redshifts. We show the power spectra of the foreground lines before and after applying a flux cut, which can suppress the foreground contamination about one order of magnitude below the signal. Note that there are several  foreground lines that contaminate the [OII] and H$\beta$ lines, so the flux cuts are for the total foregrounds. The flux cuts for each line at different redshifts within $0.8\le z\le5.2$ are shown in Table \ref{tab:SNR_fc}. We find the stronger lines, e.g. [OIII] and [OII], have higher flux cuts than the weaker line H$\beta$ at the same redshift, and the flux cuts decrease as the redshift increases, as expected. For the [OIII] and [OII] lines, the flux cuts are similar at the same redshift, which varies from $\sim10^{-22}$ to $\sim10^{-24}$ $\rm W/m^2$,  while the H$\beta$ flux cuts are about a factor of 5 lower than that of [OIII] and [OII] at $z\lesssim3$, and they become more and more similar at $z>3$. 

\subsection{Detectability}

Next, we discuss the detectability of the four lines at $0.8\le z\le\ 5.2$ with the SPHEREx experiment. First, we need to estimate the variance of the intensity power spectrum. In a line intensity survey, the variance of the power spectrum at a given redshift is given by \cite[e.g.][]{Lidz11,Gong12,Uzgil14}
\be \label{eq:DPk}
\Delta P_{\rm line}(k)^2 = \frac{\left[ P_{\rm line}(k)+{P}_N^{\rm line}(k)\right]^2}{N_m(k)},
\ee
where $P_{\rm line}$ is the line intensity power spectrum we derived in the last section that denotes the cosmic variance term, and $P_N^{\rm line}(k)$ is the noise power spectrum that depends on the instrument and takes the form
\be
P_N^{\rm line}(k)=\frac{V_{\rm pix}\,\sigma_{\rm pix}^2}{t_{\rm pix}}.
\ee
Here $\sigma_{\rm pix}^2/t_{\rm pix}$ denotes the squared instrument thermal noise per survey pixel, where $t_{\rm pix}$ is the integration time per pixel, and $V_{\rm pix}$ is the pixel volume. $N_m(k)$ is the number of Fourier modes in an interval $\Delta k$ at $k$ in the upper-half wavenumber plane. In principle, it can be evaluated by
\be \label{eq:Nm}
N_m(k) = 2\pi k^2\Delta k\,\frac{V_{\rm S}}{(2\pi)^3},
\ee
where $V_{\rm S}$ is the total survey volume. Note that there can be large discrepancy between the real $N_m$ and the estimated one from Equation (\ref{eq:Nm}), especially at large $k$. In our calculation, we count the modes explicitly to determine $N_m$ in each $\Delta k$. The signal to noise ratio (SNR) of the intensity power spectrum then can be derived by
\be \label{eq:SNR}
{\rm SNR} = \sqrt{\sum_{k\,\rm bin}\left[\frac{P_{\rm line}(k)}{\Delta P_{\rm line}(k)}\right]^2}.
\ee

In order to investigate the detectability of the intensity power spectra of the four lines, we take the SPHEREx experiment as an example to estimate the error and SNR. SPHEREx is a proposed space telescope with a diameter of 20 cm, and has four bands which cover 0.75-1.32, 1.32-2.34, 2.34-4.12 and 4.12-4.83 $\mu$m, respectively. The frequency resolution of the first three bands is $R$=41.5, and $R$=150 in the fourth band. In our study, only the first three bands are used in the discussion for $z<5$. We assume a deep SPHEREx survey with a total survey area of 200 deg$^2$ and a beam size 6.2$\times$6.2 arcsec$^2$, which has repeated observations and can provide a dataset ideal for intensity mapping \citep{Dore15}.

In Figure \ref{fig:PS_com}, we show the estimated detection errors and SNR for the clustering power spectra of the H$\alpha$, [OIII], [OII] and H$\beta$ lines at $z=1\pm0.2$. We find the SNR of H$\alpha$, [OIII], [OII] lines are SNR$>$18, which can be easily measured by  SPHEREx, but the SNR of H$\beta$ is low (SNR$\simeq$3) and difficult to detect. In Table \ref{tab:SNR_fc}, we list the SNR for each line at different redshifts. We can see that the H$\alpha$ has SNR$\simeq$7 even at $z\sim4$, and the SNRs of [OIII] and [OII] lines also are as high as SNR$\simeq$12 and 8 at $z\sim3$, respectively. This indicates that SPHEREx can provide precision measurements of the intensity power spectra of H$\alpha$, [OIII] and [OII] lines at $z\lesssim$3. However, the SNR of the H$\beta$ line is less than 3 at all redshifts, which is challenging to measure accurately. 

In Figure \ref{fig:N_f}, we show the total number of sources for each line, whose flux $f_{\rm line}$ is greater than a given value, in a survey voxel of SPHEREx at $z=1.0\pm0.2$. In Table \ref{tab:SNR_fc}, we list the percent of voxels removed above these flux cuts for each line at different redshifts. We find there are about 3\% voxels for [OIII] and [OII] line at $z=1.0\pm0.2$ that need to be masked, and it is about 12\% for H$\beta$ line. The percent of removed voxels increase quickly as the redshift increases, and about 55\%, 67\% and 87\% for [OIII], [OII] and H$\beta$ at $z=4.8\pm0.4$, respectively. For H$\beta$ line, we find the number of H$\beta$ sources is comparable to that of foreground contaminates, even after performing flux cut. This indicates that it is quite challenging to measure H$\beta$ fluctuations with SPHEREx. For [OIII] and [OII] lines in the range $z\lesssim3$ with relatively high SNR$>$3, the masked voxels are less than 10\% and 30\% for [OIII] and [OII] lines, respectively. This implies SPHEREx is capable of precisely measuring the intensity power spectra of [OIII] and [OII] lines at $z\lesssim3$ with relatively small masked voxel fraction. Of course, the H$\alpha$ line is the best observable with the strongest intensity and has no considerable foreground line contamination.

The contamination of continuum emission from galaxies and recently proposed intrahalo light (IHL) also needs to be considered \citep{Cooray12,Zemcov14,Mitchell-Wynne15,Silva15,Yue15}. We find that the total mean intensity and power spectrum of the continuum emission can be larger than those of the four optical lines by two orders of magnitude. However, since the spectrum of the continuum emission is expected to be smooth, we can remove it in observed 3D spectral line data cube by fitting polynomials (or other forms) as a function of frequency along different line of sights \citep[see e.g.][]{Yue15}. This process does not affect the flux cuts and SNRs significantly, and we will discuss it in details in our future work with simulations. In addition, the contamination of zodiacal light should be small for the line power spectrum. This is because that the spatial distribution of zodiacal light is relatively smooth with small fluctuations \citep{Zemcov14,Arendt16}, and the 200 deg$^2$ SPHEREx deep survey is planning to observe at the north and south ecliptic poles, where the zodiacal light is much fainter than that at low ecliptic latitudes.

\section{Cross correlation with 21-cm line}

In order to remove foreground contamination, we can also cross correlate different lines at the same redshift. Since the signals at the same redshift trace the same matter distribution while the foreground lines at different redshifts do not, cross correlation can effectively reduce the foreground contamination. The direct cross correlations can be performed between two lines of H$\alpha$, [OIII], [OII] and H$\beta$. In principle, the value of the SNR for the cross power spectrum should be geometric mean of the SNRs of the auto power spectra for the two cross lines in the same intensity survey. As a result, it is convenient to predict the SNR of cross power spectra between the four lines for SPHEREx experiment. We could also consider cross correlating the intensity power spectrum with optical galaxy surveys \citep{Lidz16}. In this work, we focus on the cross correlation of the four lines probed by SPHEREx with the hydrogen hyperfine-structure 21-cm line measured by CHIME and Tianlai experiments.

At low redshifts after the epoch of reionization, neutral hydrogen mainly resides in galaxies. Similar to the emission lines discussed above at $z<5$, the 21-cm line emitted by neutral hydrogen also traces the distribution of galaxies and matter in the Universe. As a result, the 21-cm line can correlate with optical lines at the same redshift. The clustering cross power spectrum is given by
\be \label{eq:P_cross}
P_{\rm cross}^{\rm clus}(k,z) = \bar{b}_{\rm 21cm}\bar{b}_{\rm line}\bar{T}_{\rm 21cm}\bar{T}_{\rm line} \,P_{\delta \delta}(k,z),
\ee
where $\bar{b}_{\rm 21cm}$ and $ \bar{T}_{\rm 21cm}$ are the bias and mean brightness temperature of the 21-cm line, respectively \citep{Gong11a}. $\bar{T}_{\rm line}$ is the mean brightness temperature of the four optical lines converted from the mean intensity $\bar{I}_{\rm line}$ by the Rayleigh-Jeans law. The $\bar{b}_{\rm 21cm}$ is expressed by
\be
\bar{b}_{\rm 21cm}(z) = \frac{\int_{M_{\rm min}}^{M_{\rm max}} dM\frac{dn}{dM}M_{\rm HI}\,b(M,z)}{\rho_{\rm HI}},
\ee
where $M_{\rm HI}(M,z)$ is the neutral hydrogen mass given by the fitting results of simulations in \cite{Gong11b}, and $\rho_{\rm HI}=\int dM (dn/dM) M_{\rm HI}$ is the mass density of neutral hydrogen. The mean 21-cm temperature can be estimated by \citep{Chang10}
\ba
\bar{T}_{\rm 21cm}(z) &=& 248\left( \frac{\Omega_{\rm HI}}{10^{-3}}\right)\left( \frac{h}{0.73}\right)\left( \frac{1+z}{1.8}\right)^{0.5} \nonumber\\  
                               &&\left[ \frac{\Omega_M+\Omega_{\Lambda}(1+z)^{-3}}{0.37}\right]^{-0.5} \rm \mu K.
\ea
Here $\Omega_{\rm HI}=\rho_{\rm HI}/\rho_c$ where $\rho_c$ is the critical density. We estimate the cross shot-noise term by 
\ba
P^{\rm shot}_{\rm cross}(z) = && \int_{M_{\rm min}}^{M_{\rm max}} dM \frac{dn}{dM} \left[\frac{L_{\rm line}}{4\pi D_{\rm L}^2}y(z)D_{\rm A}^2\right] \nonumber\\ 
&& \times \left( \bar{T}_{\rm 21cm}\frac{M_{\rm HI}}{\rho_{\rm HI}}\right),
\ea
and the 21-cm shot-noise power spectrum is given by
\be
P^{\rm shot}_{\rm21cm}(z) = \int_{M_{\rm min}}^{M_{\rm max}} dM \frac{dn}{dM} \left( \bar{T}_{\rm 21cm}\frac{M_{\rm HI}}{\rho_{\rm HI}}\right)^2.
\ee
Then the total cross power spectrum is $P^{\rm tot}_{\rm cross} = P_{\rm cross}^{\rm clus} + P^{\rm shot}_{\rm cross}$. We also calculate the cross correlation coefficient $r$, and find that $r$ is always greater than 0.8 and varies from $\sim0.8$ to $\sim1$ as $k$ becomes smaller.

\begin{table}[!t]
\caption{Design parameters for CHIME and Tianlai experiments we use. }
\vspace{-3mm}
\begin{center}
\begin{tabular}{  l  c  c  c }
\hline\hline
        & CHIME & Tianlai & Unit  \\
\hline \vspace{-2mm}\\
Survey area $A_s$& 10000 &10000 & $\rm deg^2$ \\ 
Total int. time $t_{\rm tot}$& 10$^4$ & 10$^4$  & hour\\
Total bandwidth  & 400-800 & 400-1420 & MHz \\
Redshift range & 0.8-2.5 & 0.0-2.5 & - \\
Sys. Temp. $T_{\rm sys}$ & 50 & 50 & K \\
FoV N-S & $\sim$150 & $\sim$150 & deg \\
FoV E-W  & $\sim$1.9 & $\sim$1.6 & deg \\
Cylinder size  & 100$\times$20 & 120$\times$15 &  $\rm m^2$\\
Num. of cylinders  & 5 & 8 &  -\\
Tot. Collecting area  & $10000$ & $14400$ & $\rm m^2$ \\
Num. of feeds  & 256$\times$5 & 275$\times$8 & - \\
Freq. resolution $\Delta\nu$  & $\sim$1 & $\sim$0.1 & MHz \\ \vspace{-2mm}\\
 \hline
\end{tabular}
\end{center}
\vspace{-4mm}
\begin{center}
\end{center}
\label{tab:design_para}
\end{table}

\begin{figure*}[t]
\epsscale{1.9}
\centerline{
\resizebox{!}{!}{\includegraphics[scale=0.42]{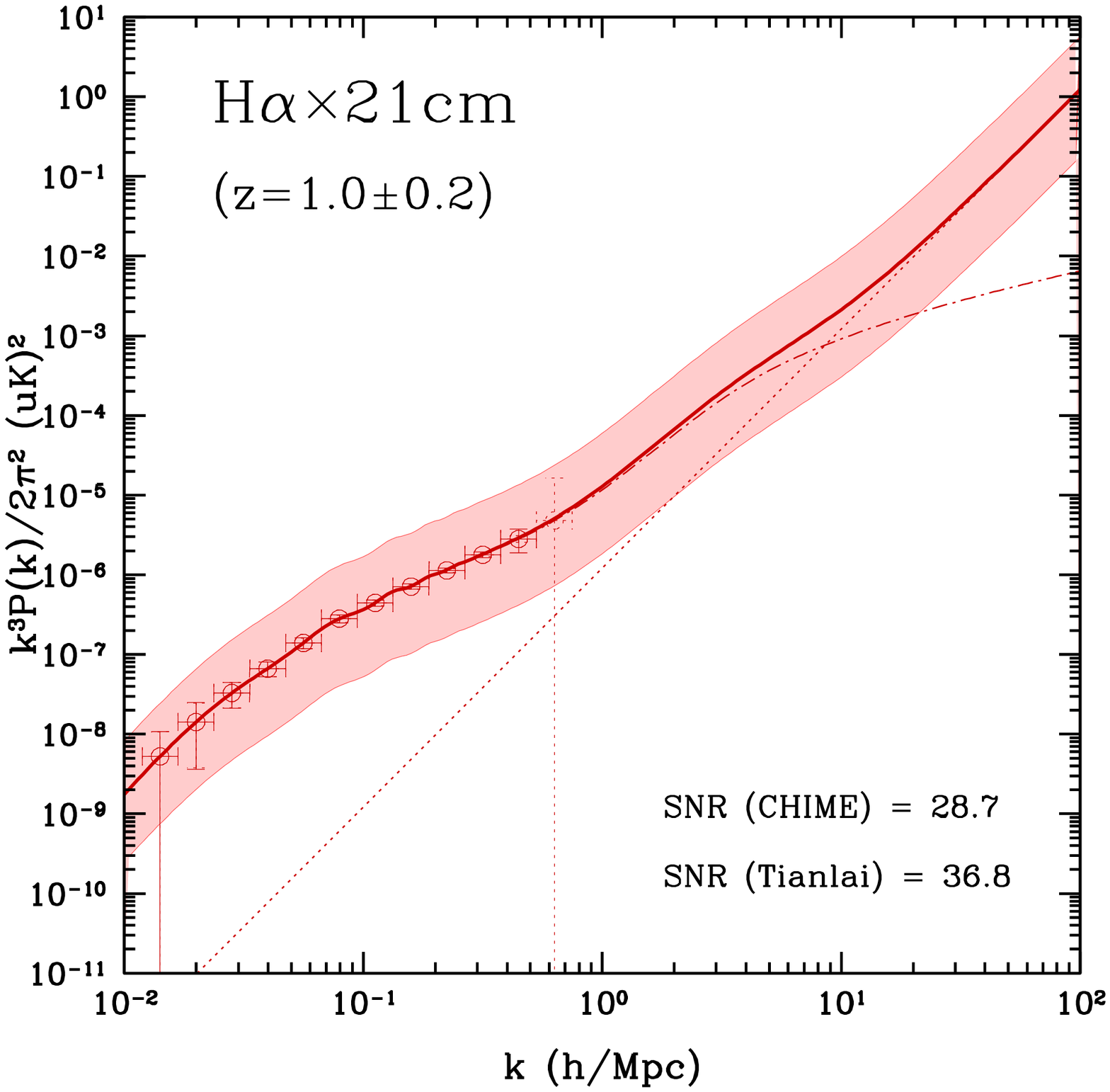}}
\resizebox{!}{!}{\includegraphics[scale=0.42]{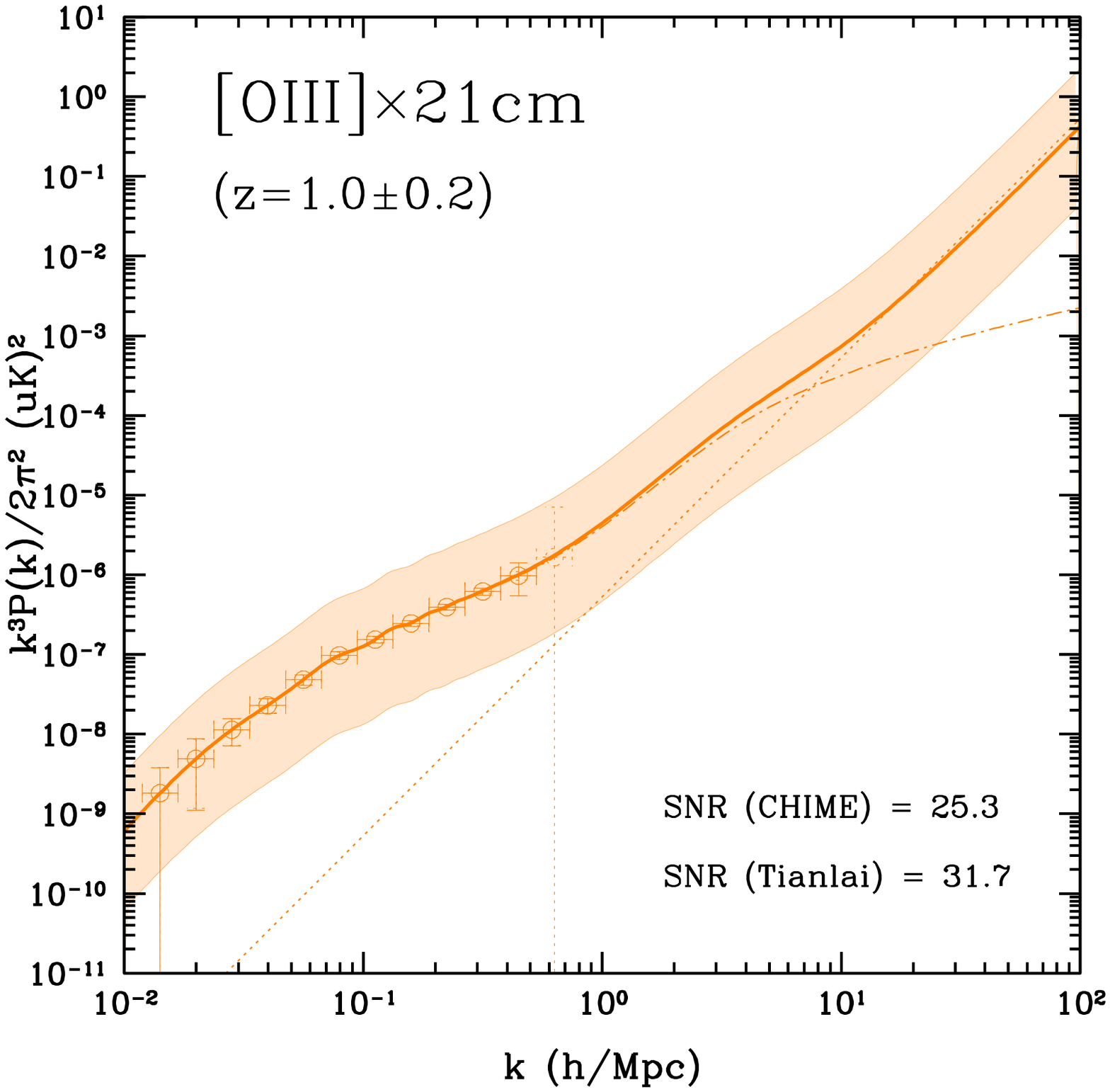}}
}
\centerline{
\resizebox{!}{!}{\includegraphics[scale=0.42]{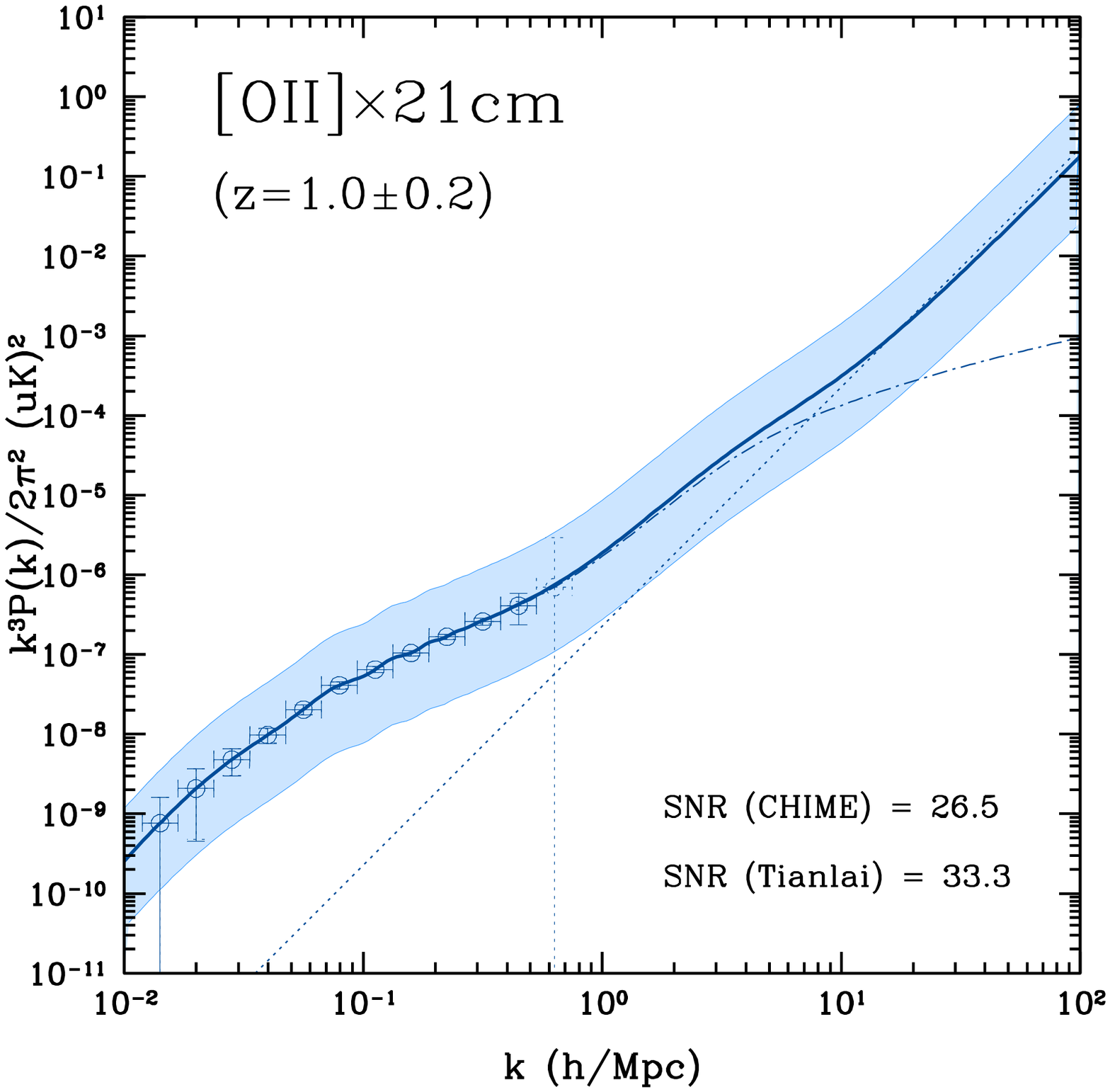}}
\resizebox{!}{!}{\includegraphics[scale=0.42]{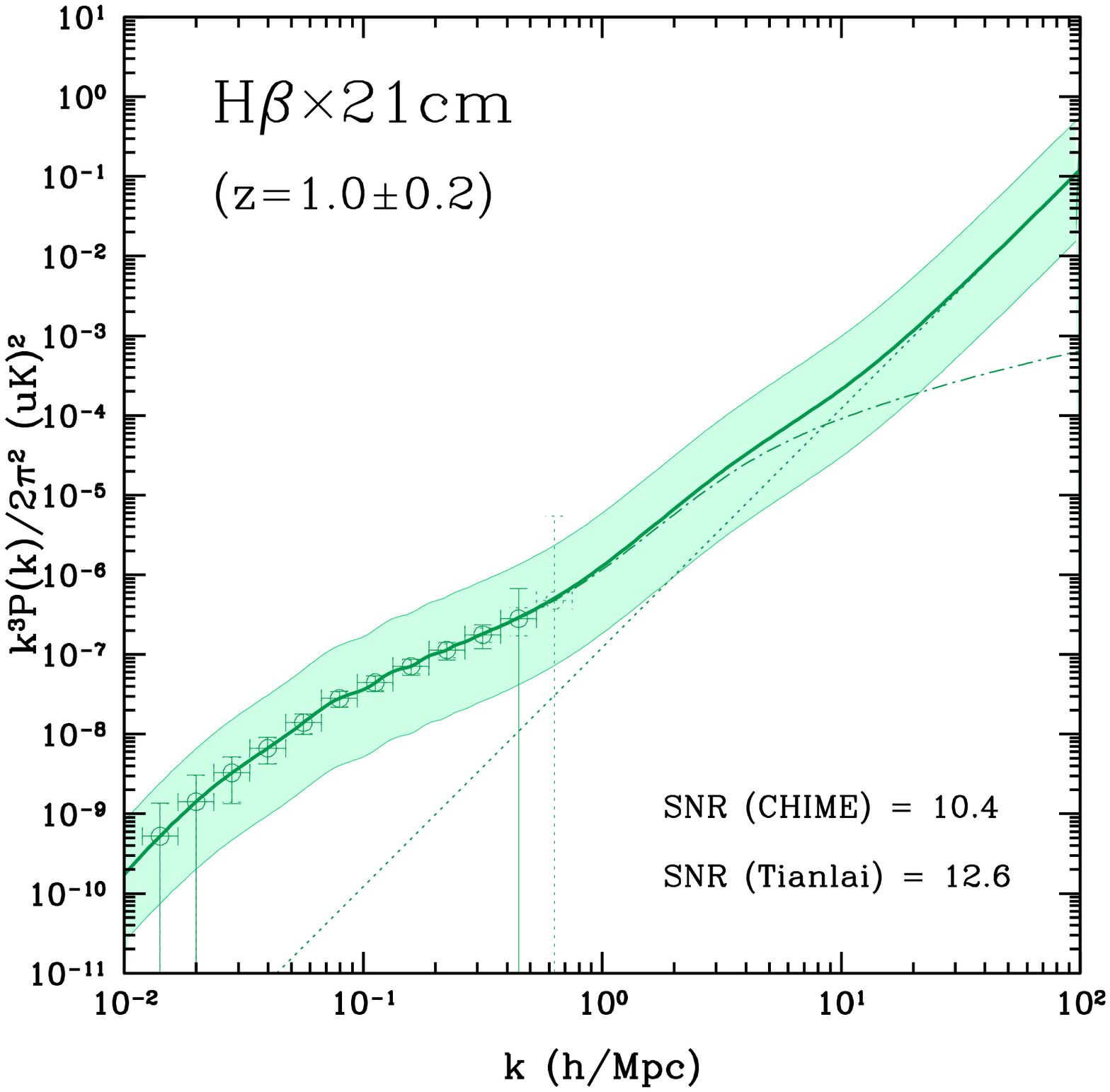}}
}
\epsscale{1.0}
\caption{\label{fig:PS_cross} The cross power spectra of H$\alpha$, [OIII], [OII] and H$\beta$ lines with the 21-cm line at $z=1.0\pm0.2$.  The errors for the CHIME and Tianlai experiments are shown as solid and dotted bars, respectively. The dashed-dotted and dotted curves denote the clustering and shot-noise terms of the cross power spectra, respectively. We find Tianlai has smaller errors and higher SNR than CHIME at a given redshift. Unlike the intensity power spectra probed by the SPHEREx,  the 21-cm experiments are restricted by the relatively low spatial and spectral resolutions, and the cross power spectra can only be measured at large scales with $k<1\ {\rm Mpc^{-1}}h$.}
\end{figure*}

In order to explore the detectability of the cross power spectrum, we also need to estimate the noise power spectrum given by the instruments, and we take the CHIME and Tianlai 21-cm experiments as examples. The CHIME and Tianlai experiments are radio interferometers with parabolic cylinder reflectors. The goal of CHIME and Tianlai is to measure the baryon acoustic oscillation (BAO) in the large scale structure of the Universe as traced by the 21-cm line from galaxies, and from such measurements to extract the properties of dark energy. Both of the experiments cover large sky fractions and redshift ranges to obtain accurate measurements of the 21-cm power spectrum and BAO features. The design parameters of these instruments can be found in Table \ref{tab:design_para}. 

We estimate the variance of the cross noise power spectrum at a given redshift by \citep[e.g.][]{Gong12,Lidz16}
\be \label{eq:P_N_cross}
\left(\Delta P_{\rm cross}\right)^2 = \frac{P_{\rm cross}^2 + (P_{\rm line}+P_N^{\rm line})(P_{\rm 21cm}+P_N^{\rm 21cm})}{2\,N_m^{\rm cross}(k)},
\ee
where $P_{\rm cross}$, $P_{\rm line}$ and $P_{\rm 21cm}$ are the intensity power spectra of cross, the four lines, and 21-cm, respectively. The cross power spectrum can arise from both clustering and total power spectra depending on the regime of interest. The $N_m^{\rm cross}(k)$ is the number of modes in a $k$ bin, and we count the modes explicitly to derive it from the smaller survey volume and larger voxel in the two different surveys. $P_N^{\rm 21cm}$ is the noise power spectrum of the 21-cm line, which can be derived from the CHIME and Tianlai experiments, and it is given by \citep{McQuinn06}
\be
P_N^{\rm 21cm} = r_c^2\,y(z) \frac{\lambda^2\, T_{\rm sys}^2}{A_e\,t_{k}}.
\ee
Here $r_c$ is the comoving distance, $A_e$ is the effective area of an antenna, and $T_{\rm sys}$ is the system temperature. The $t_{k} = t_0(A_e/\lambda^2)n({\mu_{\perp})}$ is the average observation time for a mode $k$, where $t_0$ is the total integration time and $n(\mu_{\perp})$ is the number density of the baselines. $\mu_{\perp} = k\,{\rm sin(\theta)} r_c/2\pi$, and $\theta$ is the angle between k mode of interest and the line of sight. Then the SNR of the cross power spectrum becomes similar to Equation (\ref{eq:SNR}) that needs to be replaced by the terms $P_{\rm cross}(k)$ and $\Delta P_{\rm cross}(k)$.

\begin{table*}[!t]
\caption{The SNRs for the cross correlations of H$\alpha$, [OIII], [OII] and H$\beta$ by SPHEREx with 21-cm line by CHIME and Tianlai at $0.8\le z\le2.4$.}
\vspace{-3mm}
\begin{center}
\begin{tabular}{  l  l  c  c   c   c }
\hline\hline
 Line        & 21-cm expt. & $z=1.0\pm0.2$ & $z=1.4\pm0.2$ & $z=1.8\pm0.2$ & $z=2.2\pm0.2$  \\
\hline \vspace{-2mm}\\
H$\alpha$&$\times$CHIME & 28.7 & 26.9 & 18.9 & 15.2  \\ 
                &$\times$Tianlai & 36.8 & 36.1 & 26.6 & 22.0  \\
                
 \hline   \vspace{-2mm}\\
{\rm [OIII]}&$\times$CHIME& 25.3 & 16.7 & 17.8 & 12.2  \\              
                &$\times$Tianlai & 31.7 & 21.6 & 18.4 & 17.4  \\
 \hline   \vspace{-2mm}\\
{\rm [OII]}&$\times$CHIME& 26.5 & 17.2 & 11.7 & 8.8  \\
                &$\times$Tianlai & 33.3 & 22.2 & 15.8 & 12.4   \\
 \hline   \vspace{-2mm}\\
H$\beta$ &$\times$CHIME& 10.4 & 6.8 & 7.8 & 5.2  \\ 
                &$\times$Tianlai & 12.6 & 8.5 & 10.5 & 7.3  \\ \vspace{-2mm}\\
\hline
\end{tabular}
\end{center}
\vspace{-4mm}
\begin{center}
\end{center}
\label{tab:SNR_cross}
\end{table*}

In Figure \ref{fig:PS_cross}, we show the cross power spectra of H$\alpha$, [OIII], [OII] and H$\beta$ lines with the 21-cm line at $z=1.0\pm0.2$. The errors are also shown for the CHIME and Tianlai experiments in solid and dotted bars, respectively. We find both the SNRs of the cross power spectra for the CHIME and Tianlai are large enough to be well measured for all of the four lines at $z\sim1$. The SNR of Tianlai is larger than that of CHIME, since it has greater collecting area, more receivers, and higher resolution, as shown in Table \ref{tab:design_para}. Note that the measurable scales of the cross power spectrum are larger than the intensity power spectrum probed by SPHEREx alone (see Figure \ref{fig:PS_com}). This is because the spatial and spectral resolutions of CHIME and Tianlai are relatively low, so that only large scales with $k<1\ {\rm Mpc^{-1}}h$ can be detected. We find that the detectability of SPHEREx cross-correlated with CHIME or Tianlai is greatly improved compared to the current galaxy$\times$21-cm measurements given by \cite{Chang10} and \cite{Masui13}. The SNR of SPHEREx$\times$CHIME and Tianlai is $\sim$30 for H$\alpha$, [OIII] and [OII] lines, and it is $\sim$10 for H$\beta$ line at $z\sim1$. For comparison, the SNR is less than 10 for the galaxy$\times$21-cm measurements at $z\sim0.8$.

In Table \ref{tab:SNR_cross}, we tabulate the SNRs of the cross power spectrum for the four optical lines with CHIME and Tianlai at $0.8\le z\le2.4$. As can be seen, the cross power spectra have large SNRs over the redshift range, even for the relatively faint H$\beta$ line with SNR$>$5, and the foreground line contamination can be reduced significantly by cross correlation. This indicates that cross correlations of the four lines with the 21-cm line are an advantageous method for extracting the intensity fluctuation signal.

Another similar 21-cm experiment,  Hydrogen Intensity and Real-time Analysis eXperiment (HIRAX), focuses on the similar redshift range of $0.8<z<2.5$ as CHIME and Tianlai for measuring BAO and constraining dark energy \citep{Newburgh16}. HIRAX is a new radio interferometer under development in South Africa, which is comprised of 1024 six meter parabolic dishes with frequency coverage of 400-800 MHz. It plans to observe 15000 deg$^2$ with drift-scan mode in the Southern Hemisphere in four years. HIRAX can complement the observations of CHIME and Tianlai in the Northern Hemisphere, and provide additional measurements for BAO and LSS. The detectability of the cross power spectrum with HIRAX is similar to CHIME and Tianlai.

Besides, we also estimate the detectability of cross power spectrum between SPHEREx and Square Kilometer Array phase one mid-frequency dish array (SKA1-mid)\footnote{\tt https://www.skatelescope.org/}. SKA1-mid contains one hundred and ninety 15m dishes and 64 MeerKAT dishes. Since SKA1-mid has relatively few short baselines, it has been proposed to use as a collection of single dishes for large-scale intensity mapping survey \citep[e.g.][]{Bull15}. We adopt this strategy, and assume a system temperature $T_{\rm sys}=25$ K and total integration time $t_{\rm tot}=10^4$ hours for a total survey area $A_s=10000$ deg$^2$. A frequency resolution $\Delta \nu=3.9$ kHz is adopted for SKA1-mid band one. We find the SNRs of the cross power spectra between SPHEREx and SKA1-mid are lower than that for CHIME and Tianlai. For instance, SNR=5.7 for H${\alpha}\times 21$ cm at $z=1$, which is a factor of 5$\sim$6 lower than CHIME and Tianlai. This is basically due to relatively low spatial resolution for SKA1-mid single dishes.

We need to note that the sensitivity estimates of cross power spectra above are assuming perfect foreground subtraction. Imperfect subtraction, which can be caused by imperfect instrument modeling, will result in residual foregrounds as a noise in the observational data. This can significantly affect the measurements of cross power spectra, especially when considerable residual foregrounds are left in the data.

\section{SFRD Constraints}

\begin{figure}[t]
\includegraphics[scale = 0.44]{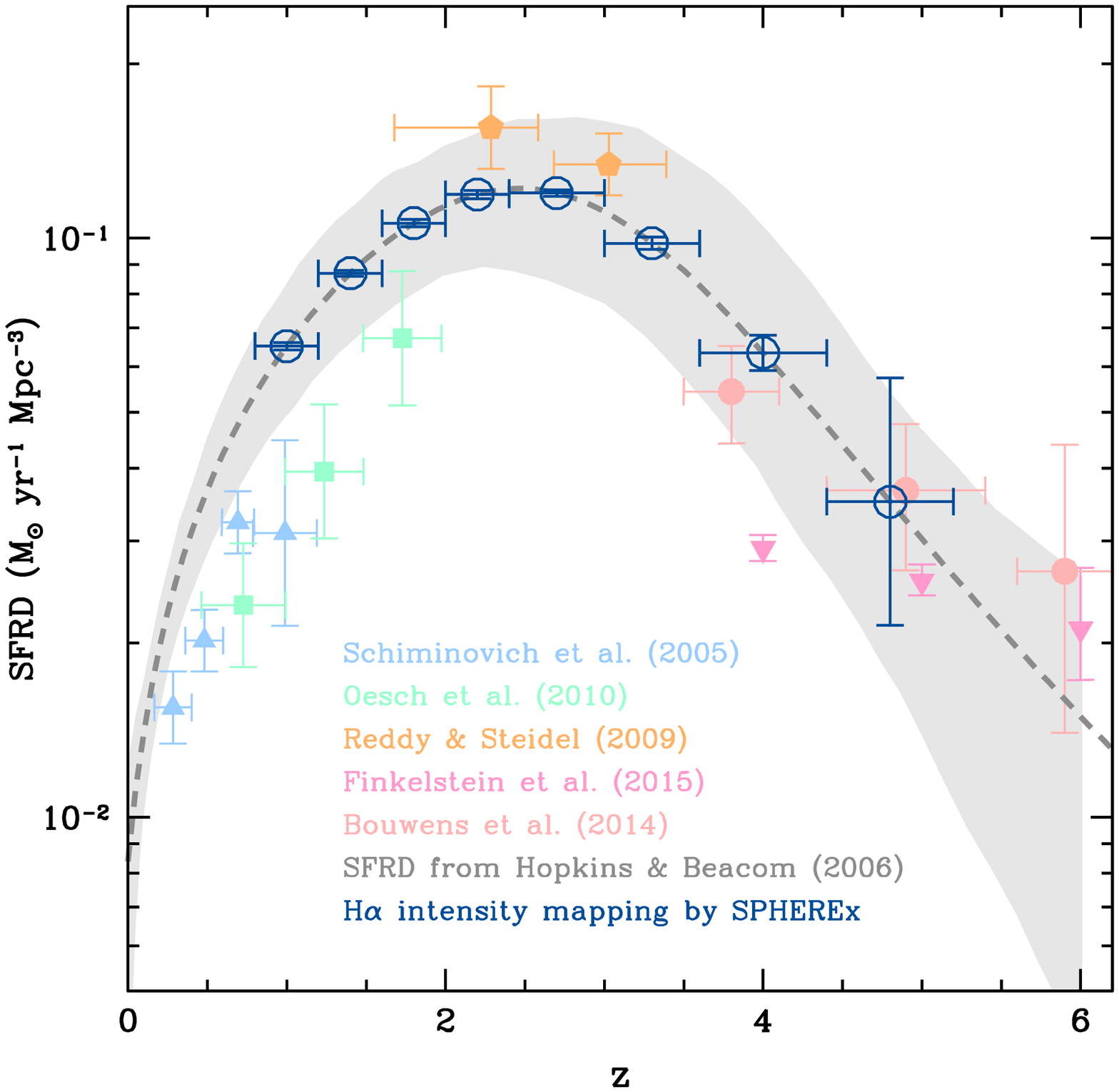}
\caption{\label{fig:SFRD_err} The constraints on the SFRD from the H$\alpha$ intensity as measured by SPHEREx. The filled blue triangles, green squares, orange pentagons, red circles and pink inverse triangles are the observational data given in other works \citep{Schiminovich05,Oesch10,Reddy09,Bouwens14,Finkelstein15}. The gray curves show the SFRD given in \cite{Hopkins06}, and the gray region shows the 2$\sigma$ C.L. We find that intensity mapping can provide stringent constraints on the SFRD at $z\lesssim 4$.}
\end{figure}

The purpose of performing intensity mapping surveys is to illustrate the galaxy distribution, measure the BAO in the large scale structure, and derive the statistical properties of galaxies and the Universe. Some important quantities can be explored by intensity mapping, such as the SFRD$(z)$, and cosmological parameters including $\Omega_M$, $\Omega_{\Lambda}$, dark energy equation of state $w$, $\sigma_8$, Hubble parameter $H$, and so on. In this section, using the Fisher matrix, we estimate the constraints on the SFRD available from emission line intensity mapping with SPHEREx. Accurate measurement of the SFRD is essential for the studies of galaxy evolution, extragalactic background light and other related fields, especially when including faint galaxies. Unlike the ordinary LF surveys by observing bright galaxies, intensity mapping could capture the emissions from faint galaxies, and provide more reliable constraints on the cosmic star formation history.

The Fisher matrix of the intensity mapping power spectrum $P(k)$ for two parameters $q_i$ and $q_j$ can be written as
\be
F_{ij} = \sum_{k\, {\rm bin}} \frac{1}{\Delta P(k)^2} \frac{\partial P(k)}{\partial q_i} \frac{\partial P(k)}{\partial q_j},
\ee
where $\Delta P(k)^2$ is the variance of the power spectrum. After obtaining the Fisher matrix, we can derive the covariance matrix of the parameters from $C_{ij}=(F^{-1})_{ij}$. For parameter errors without correlation, the covariance matrix is diagonal, and we can derive the error as $\sigma_i=\sqrt{F_{ii}^{-1}}$. In order to calculate the errors of the SFRD at different redshifts $\sigma_{\rm SFRD}(z)$, we set $q_{i,j}={\rm SFRD}$, and estimate ${\partial P(k)}/{\partial q}$ in each $k$ bin at a given redshift. The variance $\Delta P(k)^2$ can be obtained by Equation (\ref{eq:DPk}).

In Figure \ref{fig:SFRD_err}, we show the errors of the SFRD measured by SPHEREx, as estimated by the Fisher matrix. This result is derived from the H$\alpha$ clustering power spectrum over $0.8\le z\le5.2$. We take the SFRD given in \cite{Hopkins06} as the fiducial model. We find that the SFRD can be well constrained with the accuracy higher than 7\% at $z\lesssim4$. The error rises quickly at higher redshifts, since the SNR of the intensity power spectrum deceases significantly at high-$z$. For the other lines, i.e. [OIII], [OII] and H$\beta$, the constraints are not as good as for the H$\alpha$ line, especially for H$\beta$ line. However, the [OIII] and [OII] still can constrain the SFRD with good accuracy ($<6\%$) at $z\lesssim3$. These constraints are better than the ordinary method by measuring the LFs from individual galaxies. We can also make use of the cross power spectrum to measure the SFRD, which can suppress foreground line contamination. Intensity mapping surveys therefore offer an efficient way to probe the SFRD at different redshifts. 

We notice that the Fisher matrix estimation actually produces the smallest errors for the parameters, which assumes Gaussian probability distribution. There is also degeneracy between the SFRD and other parameters, such as galaxy bias, which could enhance the uncertainty of the SFRD constraint. A more accurate and reliable method is to use the Markov Chain Monte Carlo (MCMC) method to estimate the errors and include degeneracies between parameters. Besides the SFRD, we could also investigate constraints on cosmological parameters available from intensity power spectrum studies. We will discuss these in our future work.

\section{Summary and discussion}

In this work, we investigated intensity mapping of the H$\alpha$, [OIII], [OII] and H$\beta$ lines at $0.8\le z\le5.2$. We first estimated the mean intensities of the four optical emission lines at different redshifts using three methods, i.e. the observed LFs, simulations of the galaxy SFR, and SFRD$(z)$ derived from observations. We find the results of the three methods are consistent with one another in 1$\sigma$ C.L. for all four lines. We also have taken account of dust extinction in estimates, so that the mean intensities we obtain are the observed intensities.

Besides the mean intensity, the fluctuation of the intensity is the main focus of our study. We calculate the intensity power spectra for the four lines using halo model at $0.8\le z\le5.2$. We find that the power spectra have the similar amplitudes at $1\lesssim z\lesssim2$ due to the increasing SFRD and bias over this redshift range. The intensity power spectra drops significantly at $z\gtrsim3$. This implies that it is challeging to measure the intensity power spectrum at high redshifts, although we can make strong detections at $z<3$.

Foreground line contamination is important to intensity mapping surveys. We assume there is no large foreground contamination for the H$\alpha$ line, though [OIII], [OII], and especially H$\beta$ are more challenging. We explored flux cuts that can suppress foregrounds by approximately one order of magnitude below the signal for the [OIII], [OII] and H$\beta$ lines. The [OIII] and [OII] have similar flux cut thresholds at low redshifts, and both are larger than those of H$\beta$ line. 

In order to study the detectability of the intensity power spectra of the four lines, we take the proposed SPHEREx experiment as an example to evaluate errors and SNRs. As expected, H$\alpha$ has the highest SNRs at all redshifts with SNR$>$7 at $z\lesssim$4. The SNRs of the [OIII] and [OII] are also as high as SNR$>$8 at $z<3$, and $<$3 for the H$\beta$ line over the redshift range of interest. We also estimate the percentage of the SPHEREx survey voxels that need to be masked to suppress foregrounds. We find this percentage is less than $\sim$10\% and 30\% for the [OIII] and [OII], respectively, at $z<3$, which indicates that it is feasible to mitigate the foregrounds for the [OIII] and [OII] lines. 

Another method to reduce the foreground contamination is to cross correlate two lines at the same redshift. We investigate the cross correlations of 21-cm line measured by the CHIME and Tianlai experiments with the four optical lines probed by SPHEREx at $0.8\le z\le 2.4$. We find the SNRs of the cross power spectra are large enough to detect all four lines, even for relatively faint H$\beta$ (which gives SNR$>5$ at $z<2.4$). This suggests that the cross correlation of optical lines with the 21-cm line provides a reliable way to extract the signals.

Finally, we predict the constraints on the SFRD$(z)$ from the intensity mapping of the H$\alpha$, [OIII], [OII] and H$\beta$ lines with SPHEREx. The Fisher matrix is used to generate our prediction, and we find the intensity mapping can provide a stringent constraint on the SFRD at $z\lesssim4$. The accuracy is higher than 7\% at $z\lesssim4$ for H$\alpha$ intensity mapping, and higher than 6\% for the [OIII] and [OII] at $z\lesssim3$. This constraint is tighter than the ordinary method by the LF measurements from individual galaxies.  Besides constraining the SFRD, intensity mapping also can constrain the properties of dark matter, dark energy, and cosmological parameters, since it captures the evolution of the large scale structure of the Universe. We expect that intensity mapping will have an increasingly important role in studies of galaxy evolution and cosmology in the future.

\begin{acknowledgments}
YG acknowledges the support of Bairen program from the National Astronomical Observatories, Chinese Academy of Sciences. YG thanks Qi Guo for helpful discussion about the simulation results. YG and AC acknowledge the supports from NSF CAREER AST-0645427 and AST-1313319, and the NASA grants NNX16AF39G and NNX16AF38G. MBS thanks the Netherlands Foundation for Scientific Research support through the VICI grant 639.043.006. CF acknowledges support from NASA grants NASA NNX16AJ69G and NASA NNX16AF39G.  M.G.S. acknowledges support from South African Square Kilometre Array Project and National Research Foundation. XLC acknowledges the support of the MoST 863 program grant 2012AA121701, the CAS Strategic Priority Research Program ``The Emergence of Cosmological Structures"  XDB09020301, and the NSFC through grant No. 11373030. Part of the research described in this paper was carried out at the Jet Propulsion Laboratory, California Institute of Technology, under a contract with the National Aeronautics and Space Administration. 
\end{acknowledgments}


\end{document}